\documentclass{article}

\usepackage[main,preprint]{neurips_2026}
\usepackage{amssymb}
\usepackage[utf8]{inputenc} 
\usepackage[T1]{fontenc}    
\usepackage{hyperref}       
\usepackage{url}            
\usepackage{booktabs}       
\usepackage{amsfonts}       
\usepackage{nicefrac}       
\usepackage{microtype}      
\usepackage{xcolor}         

\usepackage{amsmath}
 \usepackage{graphicx} 
\usepackage{comment}
\usepackage{algorithm}
\usepackage{algorithmic}
\usepackage{ulem}
\usepackage{enumitem}
\usepackage{graphicx}
\usepackage{subcaption}
\newcommand{\red}[1]{\textcolor{red}{#1}}
\newcommand{\blue}[1]{\textcolor{blue}{#1}}

\title{Hierarchical Multi-Agent Reinforcement Learning for Multi-Group Tax Game}

\author{%
  Honglei Guo \\
  College of Artificial Intelligence\\
  Zhejiang University\\
  State Key Laboratory of General Artificial Intelligence, BIGAI\\
  \texttt{guohonglei@bigai.ai} \\
  \And
  Yexin Li \\
  State Key Laboratory of General Artificial Intelligence, BIGAI\\
  \texttt{liyexin@bigai.ai} \\
    \And
  Chiyuan Wang \\
  Peking University\\
  State Key Laboratory of General Artificial Intelligence, BIGAI\\
  \texttt{wangchiyuan@bigai.ai} \\
    \And
  Yuhan Zhao \thanks{Corresponding author.} \\
  State Key Laboratory of General Artificial Intelligence, BIGAI\\
  \texttt{wangchiyuan@bigai.ai} \\
}

\begin{document}

\maketitle

\begin{abstract}

Taxation is a fundamental instrument of economic policy and has long been studied through economic models. However, most existing taxation models focus on interactions within a single economic group, typically with one government and multiple households. Such a setting overlooks interactions among independent economic groups, where governments may compete for economic resources through fiscal policies.
To capture these interactions, we formulate taxation as a hierarchical multi-group tax game. Within each group, a government sets tax policies while households respond, forming a hierarchical government--household interaction. Across groups, governments compete through fiscal policies, coupling the economic dynamics of distinct groups. 
This coupled structure poses challenges for standard multi-agent reinforcement learning (MARL) methods. To this end, we propose Multi-Group PPO (MGPPO), a bilevel MARL algorithm designed for hierarchical multi-group interactions. MGPPO incorporates Hierarchical Sampling to coordinate learning across agent levels and Curriculum Learning to improve training stability. We further develop a multi-group taxation simulation environment grounded in classical economic models, supporting the evaluation of fiscal policies under inter-group competition. Simulations show that in the two-group settings, MGPPO reduces the income and wealth Gini coefficients by 5.0\% and 4.9\%, respectively, and increases mean GDP by 4.1\% compared with IPPO.


\end{abstract}

\section{Introduction} \label{sec:intro}

Taxation is a central tool for influencing economic activity and the distribution of income and wealth \citep{ikeda2002dynamics,johnson2014mixed}. Designing tax policies requires accounting for many participants, including governments, households, firms, and banks, with different information and objectives. Their interactions create feedback: tax policies affect household decisions, while aggregate household behavior shapes the economic conditions under which governments set policies. These differences and feedback effects make analytical policy optimization difficult \citep{axtell2000agents,zhang2021multi}.

Multi-agent reinforcement learning (MARL) offers a computational approach to studying these interactions. It allows agents with different observations and objectives to learn policies through repeated interactions in a simulated economy. For example, the AI Economist uses two-level reinforcement learning to learn tax policies alongside agents' economic behavior \citep{zheng2022ai}. TaxAI provides a scalable economic simulator and a bilevel mean-field learning method for interactions between a government and many households \citep{mi2023taxai} . These approaches capture how a government's tax policies and household behavior influence each other. However, their focus is a single economic group governed by one government.

In reality, economic groups also interact with one another, allowing policies in one group to affect economic outcomes in others. One form of such interaction is competition between governments, where each government must consider how rival policies affect its own economy. In this paper, we focus on competition for capital. When capital can move across groups, tax policies influence how it is allocated among them \citep{ZODROW1986356}. Each government must therefore account for both household responses within its group and the policies of competing governments when setting tax policies.

These cross-group effects introduce dependencies that the single-group approaches discussed above do not capture. Combining government–household interactions within groups with competition among governments creates a coupled learning problem. Each government must learn while both household behavior and rival governments' policies are changing. To address this problem, we introduce the Multi-Group Hierarchical Tax Game (MHTG) and an associated learning algorithm. MHTG models government–household interactions within each group (intra-group) as a leader–follower game and competition for capital across groups (inter-group) as a noncooperative game among governments. Building on this formulation, we propose Multi-Group PPO (MGPPO), a bilevel on-policy MARL algorithm for learning government and household policies across interacting groups. MGPPO uses hierarchical sampling to coordinate learning across government and household levels. It also uses a curriculum that gradually introduces inter-group competition to stabilize learning. We further develop a multi-group taxation simulation environment to evaluate the learned policies and study their economic outcomes. Our contributions are as follows: 
\begin{itemize}
    \item \textbf{Multi-group tax game formulation:} We formulate taxation as a set of government–household leader–follower games linked through competition for capital among governments.
    \item \textbf{Multi-group PPO algorithm:} We develop MGPPO to learn policies for governments and households across interacting groups. The algorithm combines hierarchical sampling with a curriculum designed to support learning under inter-group competition.
    \item \textbf{Simulation environment and evaluation:} We build a multi-group taxation environment grounded in classical economic models and evaluate policy performance, training stability, and scalability. In the two-group settings, MGPPO reduces the income and wealth Gini coefficients by 5.0\% and 4.9\%, and increases mean GDP by 4.1\% compared with IPPO.
\end{itemize}
The simulation code is available at https://github.com/zhejing672/MHTG

\section{Related Work} \label{sec:related}

\paragraph{Classic Tax Models in Economics}

Taxation has been widely studied in public finance. Early works primarily adopted a normative perspective where the government acts as a social planner and optimizes welfare under household responses. Classical optimal taxation theory models household labor supply and consumption as reaction functions to tax policies, typically within static or single-level optimization frameworks~\citep{mirrlees1971exploration,atkinson1980lectures}. While these models provide important theoretical insights, they do not explicitly account for strategic interactions among decision-making entities. Subsequent research transitioned toward game-theoretic formulations to capture the strategic interplay between governments and households, typically characterizing taxation as a leader--follower (Stackelberg) game. A significant advancement in this lineage is the integration of the Bewley--Aiyagari framework \citep{10.2307/2118417,doi:10.1086/601445}, which enables modeling the responses of heterogeneous households to tax policies under idiosyncratic risk and incomplete markets. To address fiscal competition across regions or countries, tax competition has been extensively studied where multiple governments strategically set tax rates within a non-cooperative game framework~\citep{wilson1999theories,KEEN199733}. Seminal works, such as the Zodrow--Mieszkowski \citep{ZODROW1986356} model, formalize inter-jurisdictional tax competition and highlight key economic phenomena, including fiscal externalities, tax base mobility, and strategic interdependence among governments.

\begin{figure}[] 
	\centering
	\includegraphics[width=1\textwidth]{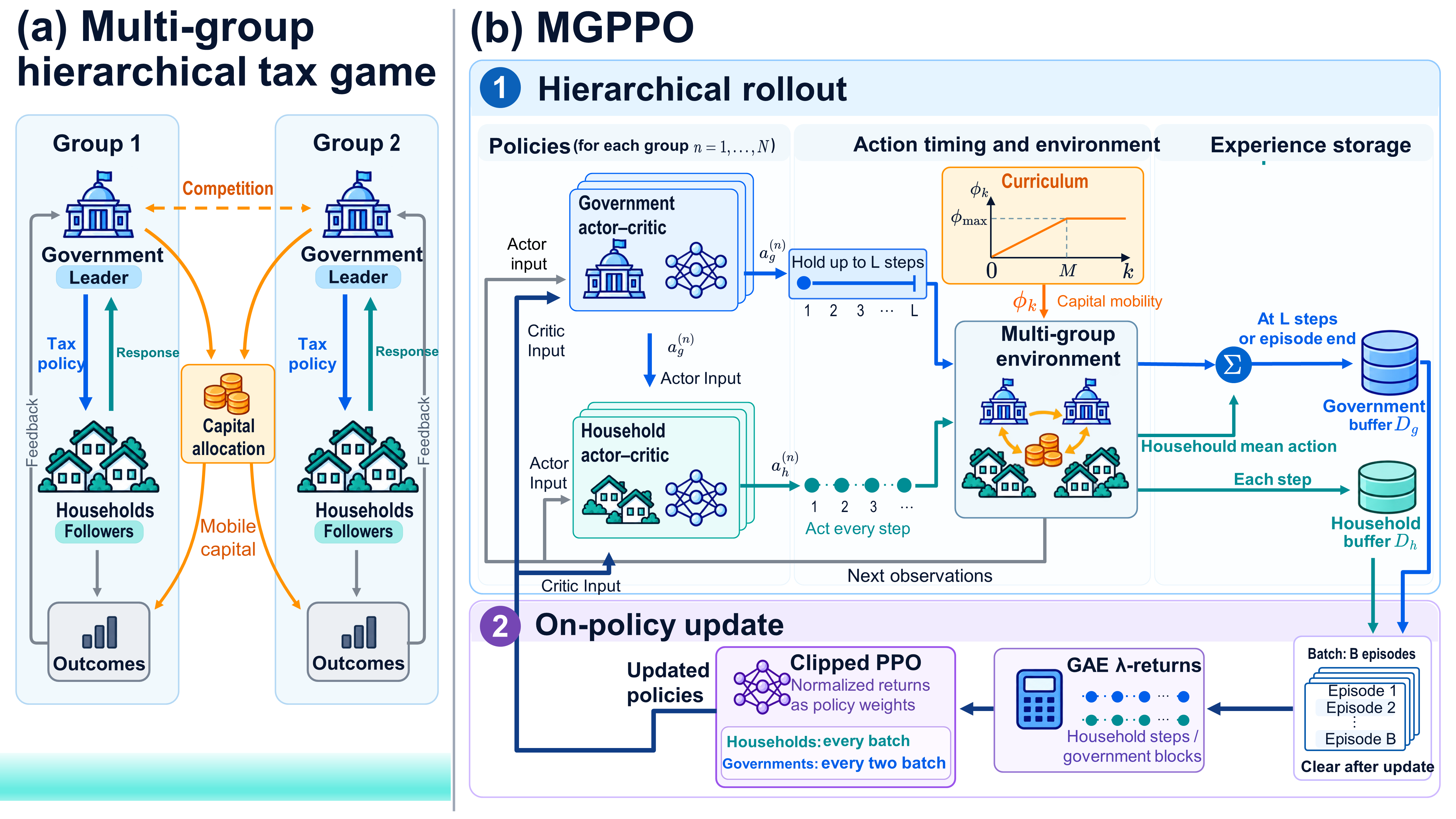}

	\caption{Framework overview. (a): Economic activities among the government, the capital allocation intermediary including firm and bank, and households. 
    (b): MGPPO workflow. Government actions are held for up to $L$ steps, households act at every step, and the curriculum parameter $\phi_k$ gradually increases inter-group coupling. Separate government and household trajectories are used for on-policy updates.
 }
	\label{fig:pipeline}
\end{figure}

\paragraph{Game Theory in Economic Analysis}

Game theory provides the fundamental mathematical formalism for analyzing strategic interactions in taxation and fiscal policy models~\citep{fudenberg1991game}. To capture the nested nature of modern taxation systems, existing studies typically employ different game-theoretic formulations to model interactions with asymmetric and symmetric strategic power. For instance, leader--follower (Stackelberg) games \citep{von2010market} are commonly adopted when capturing the asymmetric interactions between governments and households.
Here, the government acts as the leader by setting tax policies while anticipating households’ reaction functions, like adjustments in consumption and labor supply. The resulting bilevel structure provides a principled way to analyze how fiscal policies shape individual behavior~\citep{colson2007overview}. By contrast, interactions among multiple governments are often characterized by relatively symmetric strategic power and are modeled as non-cooperative games. Such characterization is widely used to study strategic competition among jurisdictions and to explain phenomena such as the failure of coordination and the emergence of competitive tax policies.
While game theory provides powerful insights, seeking equilibrium solutions for large heterogeneous agents with high-dimensional state spaces becomes analytically intractable. This necessitates the application of computational methods such as MARL.

\paragraph{Multi-Agent Reinforcement Learning in Economic System} 
MARL has developed into a broad family of algorithms for economic decision-making. Tax simulations, in particular, often involve strategic interactions among agents \citep{shi2023deep,trott2021building,hinterlang2021optimal,chen2021deep}. On-policy methods can be advantageous in such settings because they limit the mismatch between collected trajectories and current joint policies in a non-stationary game \citep{foerster2017stabilising}. Representative on-policy MARL algorithms include IPPO \cite{de2020independent}, MAPPO \citep{yu2022surprising} and HAPPO \cite{zhong2024heterogeneous}.
However, these general-purpose algorithms do not by themselves define how agents interact within a game, leaving the interaction structure to the environment and learning design.  To address this issue, some studies model these interactions explicitly  or implicitly. Explicit modeling directly influences an agent’s current decision, for example by including another agent’s action in its policy input. Implicit modeling influences decisions through the training process, such as through rewards or value estimates, and thereby changes how the agent acts in the future. As an example of explicit game modeling, Zhang et al.~\cite{zhang2020bi} formulated multi-agent coordination as a bi-level Stackelberg game and proposed an actor--critic method to learn its equilibrium.
Recently, frameworks such as the AI Economist~\citep{zheng2022ai} and the RBC-based model~\citep{curry2022analyzing} further leveraged reinforcement learning to capture bi-level interactions between policymakers and economic agents. Building on these efforts, TaxAI~\citep{mi2023taxai} extended RL-based economic simulation to large-scale heterogeneous agent settings. 
 However, these economic simulation frameworks primarily study interactions within a single government-led system. We extend this setting from one high-level agent to multiple high-level agents. Our proposed algorithm models both intra-group dynamics and strategic competition across multiple government--household systems by combining explicit and implicit modeling.

\section{Hierarchical Multi-Group Tax Game} \label{sec:game}
We propose the multi-group tax game to describe the complex phenomenon in an economic society. Figure~\ref{fig:pipeline}(a) illustrates its structure, showing hierarchical interactions among four entities: households, governments, firms, and banks.
Within each group, the government acts as the \textit{high-level agent} and optimizes social welfare (measured by GDP and the Gini coefficient) through tax policy adjustments. Households serve as \textit{low-level agents} and maximize individual utilities by choosing labor supply and consumption. Firms and banks act as \textit{intermediary institutions}  and enable the conversion and circulation of labor and capital. Across groups, governments compete for mobile capital to advance their own economic interests. Thus, we have two types of strategic interactions.
The \textit{intra-group interaction} models the government-household hierarchy via a leader--follower game grounded in the Bewley--Aiyagari framework. The \textit{inter-group interaction} captures tax competition among governments through a noncooperative game based on the Zodrow--Mieszkowski model. Full mathematical details are provided in Appendices~\ref{app:ba-model}--\ref{app:zm-dynamics}.

\subsection{Intra-Group Leader--Follower Game} \label{sec:game.intra}

Following the notations in Appendix~\ref{app:ba-model}, we introduce the actions and objectives of four entities, respectively.

\paragraph{Households}
A household has two actions: saving ratio $p_t$ and working hour $h_t$ at time $t$. We also denote $i_t$, $a_t$, and $c_t$ as the household's income, wealth, and consumption, respectively.
Under the assumption of the fixed consumption tax rate \eqref{eq:T_cons}, 
the household's consumption $c_t$ can be represented in terms of the saving ratio $p_t$ by 
\(c_t=\frac{1-p_t}{1+\tau_s}
\left[i_t-T_{\mathrm{inc}}(i_t)+\frac{G_t}{N}\right]\),
where $T_\text{inc}(\cdot)$ and $T_\text{ast}(\cdot)$ are income and asset tax functions. \(G_t\) denote total government transfers and \(N\) the number of households receiving those transfers.
Based on the working--spending dynamics in \eqref{eq:it}--\eqref{eq:at1}, each household seeks to optimize her objective function given by
\begin{equation}
	\label{eq:household_obj}
	\max_{p_t, h_t} \quad \mathbb{E} \sum_{t=0}^T \beta^t \left( \frac{(c_t)^{1-\theta}}{1-\theta} - \frac{(h_t)^{1+\gamma}}{1+\gamma}  \right),
\end{equation}
where $T > 0$ is the decision horizon and $\beta \in [0,1]$ is the discounted factor. The objective captures the household’s utility (or “happiness”), which increases with consumption $c_t$ and decreases with labor effort $h_t$. The two terms are shaped by risk preferences, with $\theta>0$ and $\gamma > 0$ representing the coefficients of relative risk aversion (CRRA) for consumption and disutility of labor, respectively.

When there are $M$ households present, they are independent and heterogeneous, distinguished by their individual productivity levels \eqref{eq:et} and financial states. Each household evolves according to the aforementioned working--spending dynamics. Although decisions are made independently, their aggregate behaviors jointly influence the wage rate and, consequently, the government’s tax policy. 
For instance, the wage rate defined in \eqref{eq:wt} is determined by the joint labor supply of all households.

\paragraph{Government} The government interacts with households through two adjustable policy instruments: tax policies and the spending ratio $\eta_t$.
Regarding taxation, we consider three types of taxes: consumption, income, and asset taxes. Based on the tax policy selection in \eqref{eq:T_cons}--\eqref{eq:T_ast}, the government chooses a fixed consumption tax rate $\tau_s$, the income tax parameter $\theta_i:=(\tau_i, \xi_i)$, and the asset tax parameter $\theta_a:=(\tau_a, \xi_a)$.
In addition to taxation, the government determines the scale of its fiscal expenditure 
to support public spending, defined by $\eta_t = \frac{G_t}{Y_t}$. Here, $G_t$ and $Y_t$ are the government's social spending and the total production. The debt dynamics in \eqref{eq:Bt} explicitly link the government’s policy choices (captured by $\theta_i, \theta_a$, and $\eta_t$) to its fiscal state trajectory.

Based on the production function \eqref{eq:Yt} and the debt dynamics \eqref{eq:Bt}, the government seeks to optimize a social welfare objective that emphasizes both economic efficiency and equality, which is 
the expected cumulative product of production and the inverse Gini index over the planning horizon:
\begin{equation}
	\label{eq:gov_obj}
	\max_{\theta_i, \theta_a, \eta_t} \quad \mathbb{E} \sum_{t=0}^T \beta^t \log(Y_t) \cdot \left(1-\operatorname{Gini}(\{i^j_{t}\}, \{a^j_{t}\}) \right)
\end{equation}
where $Y_{-1}$ is given as the initial baseline production output and $\operatorname{Gini}(\{i^j_{t}\}, \{a^j_{t}\})$ denotes the joint Gini index over all households’ income and asset distributions.
The objective balances two competing dimensions of social welfare. The term $1-\text{Gini} \in [0,1]$ measures equality, attaining 1 under perfect equality and approaching 0 under maximal concentration. The term $\log(Y_t)$ captures economic efficiency via aggregate output, where 
the logarithm is used to reduce the scale for numerical stability.

\subsection{Inter-Group Competition Game} \label{sec:game.inter}

We model inter-group interaction as a strategic competition among governments. 
This is motivated by the observation that cross-group dynamics are primarily driven by group leaders, who both interact most directly with one another and exert the greatest influence over their respective followers. 

The competition is characterized by \textit{capital regulation}. Each government can strategically set a capital tax rate $\tau^{(n)}_{t}$ to either attract or repel mobile capital and further influence its domestic capital stock and objective in \eqref{eq:gov_obj}. Let the superscript $(n)$ denote the quantities related to the $n$-th group. 
$K^{(n)}_t, B^{(n)}_t$, $M^{(n)}$ represent the capital, the debt, and the household number of the $n$-th group at time $t$, respectively.
Based on the law of motion for capital in Appendix~\ref{app:zm-dynamics}, the real capital in group $n$ is given by
\begingroup
\small
\begin{equation}
	\label{eq:combined_K}
	\begin{split}
		&K^{(n),\text{real}}_{t+1} = \underbrace{K^{(n)}_{t+1}}_{\text{Internal Accumulation}} + \underbrace{\Delta K^{(n),\text{flow}}_{t}}_{\text{External Competition}} \\
        &= (1+r_{\text{interest}}) K^{(n)}_t + (1+r_{\text{interest}}) \left( B^{n}_t - \sum_{j=1}^{M^{(n)}} a^{j,(n)}_{t} \right) + \sum_{j=1}^{M^{(n)}} a^{j,(n)}_{t+1} \quad - B^{(n)}_{t+1} - \phi \left(\tau^{(n)}_{t} - \bar{\tau}_{t}\right) K^{(n)}_{t},   
	\end{split}
\end{equation}
\endgroup
where $\bar{\tau}_t$ is the average capital tax rate across all groups at time $t$. All other variables are extended to the $n$-th group analogously and retain their original economic interpretations.

\section{Multi-Group Bilevel Learning Framework} \label{sec:famework}

\subsection{Environment Setting} \label{sec:framework.setting}
To interface with standard MARL methods, we express the model in the form of a partially observable Markov game, defined by the tuple
$\langle \mathcal{N}, \mathcal{S}, \mathcal{O}, \mathcal{A}, \mathcal{P}, \mathcal{R}, \beta \rangle$.
This representation is adopted for notational compatibility rather than as an additional modeling assumption.
Here, $\mathcal{N}$ denotes the set of agents and $\mathcal{S}$ the state space. Each agent $i \in \mathcal{N}$ has observation and action spaces $\mathcal{O}_i$ and $\mathcal{A}_i$, with joint spaces $\mathcal{O} = \prod_i \mathcal{O}_i$ and $\mathcal{A} = \prod_i \mathcal{A}_i$. The reward function is $\mathcal{R} = \{R_i(s,a)\}_{i \in \mathcal{N}}$, where $R_i(s,a)$ is the reward to agent $i$ under state $s$ and joint action $a$. The transition kernel $\mathcal{P}$ governs the system dynamics, and $\beta \in [0,1]$ is the discount factor.
We partition $|\mathcal{N}|$ agents to $N$ groups. Each group consists of one government and $N^{(n)}-1$ households, $n = 1, \dots, N$. Since all governments share a common observation and action space, and similarly for households, we omit group indices when defining these spaces for simplicity.

\paragraph{Action space} A government agent's action at time $t$ is $a_{\text{gov},t} := (\theta_i, \theta_a, \eta_t, \tau_t)$, which contains tax parameters, social spending ratio, and capital tax rate. Each household $j$ in group $n$ takes action $a^j_{\text{hh},t} := (p^j_{t}, h^j_{t})$, representing its saving ratio and working hours.

\paragraph{Observation space} In practice, it is unrealistic to assume that a government can observe all individual household states. Following \cite{mi2023taxai}, we approximate household-level information using aggregate statistics. Specifically, at time $t$, households are partitioned into the top 10\% richest and 50\% poorest, and we compute their averaged income $(\bar{i}^{\text{top}}_t, \bar{i}^{\text{btm}}_t)$, averaged asset $(\bar{a}^{\text{top}}_t, \bar{a}^{\text{btm}}_t)$, and averaged productivity $(\bar{e}^{\text{top}}_t, \bar{e}^{\text{btm}}_t)$. These aggregates, together with the current wage rate $w_t$, form the core of  the government’s observation.
Moreover, due to inter-group competition, each government also observes the previous-period government actions, denoted by ${a}^{{n}}_{\text{gov}, t-1}$.
The observation for a government in group $n$ at time $t$ is
$o^{{n}}_{\text{gov},t} = \left( w_t^{{n}}, \bar{i}^{\text{top}{n}}_t, \bar{i}^{\text{btm},{n}}_t, \bar{a}^{\text{top},{n}}_t,\bar{a}^{\text{btm},{n}}_t, \bar{e}^{\text{top},{n}}_t,\bar{e}^{\text{btm},{n}}_t, {a}^{{n}}_{\text{gov},t-1}\right).$

Households observe both their own financial state and the same aggregated group statistics (macro-level trends). For household $j$ in group $n$, the observation is
 $o^{j,n}_{\text{hh},t} = \left( w_t^{ {n}}, i^{j,{n}}_{t}, a^{j,{n}}_{t}, e^{j,{n}}_{t}, \bar{a}^{\text{top},{n}}_t,\bar{a}^{\text{btm},{n}}_t, \bar{e}^{\text{top},{n}}_t,\bar{e}^{\text{btm},{n}}_t \right).$

\paragraph{Interactive dynamics}

At time $t$, 
within each group, the government (the leader) 
observes aggregate household statistics $o_{\text{gov},t}$ and sets policy parameters $(\theta_i, \theta_a, \eta_t)$ with fiscal dynamics governed by~\eqref{eq:Bt}. Households, as followers, observe their individual financial states $(i_t, a_t, e_t)$ together with the governments action $a_{\text{gov},t}$, and choose 
decisions $(p_t, h_t)$. 
Across groups, in addition to $o_{\text{gov},t}$, each government observes the previous-period policies of other governments $a^{(-n)}_{\text{gov},t-1}$ and adjusts its capital tax rate $\tau_t$. These decisions influence the allocation of mobile capital via the flow rule~\eqref{eq:zm_flow}, thereby affecting production and household dynamics in all groups. This coupling induces strategic interdependence among governments and links the evolution of otherwise separate groups.

\subsection{MGPPO Algorithm Design}

MGPPO algorithm consisting of inter-group and intra-group modules.  We adopt a hybrid modeling strategy that combines implicit and explicit representations to characterize the underlying hybrid game structure.  The implicit modeling mechanism is employed to capture the latent competitive relationships among different groups, where strategic interactions are reflected through learned representations. In contrast, the explicit modeling mechanism is utilized to represent the leader-follower relationships, where the hierarchical decision-making process is directly incorporated into the policy estimation.
Moreover, such a design provides a balance between centralized training and scalable optimization. On the one hand, the proposed method preserve essential global interaction information during training, thereby improving learning stability in complex multi-agent environments. On the other hand, instead of directly concatenating all agents' states and actions, the architecture distributes agents’ state and action information more evenly across the actor and critic networks by combining explicit and implicit modeling,  thereby reducing the input dimensionality of each network and alleviate the optimization difficulty caused by large-scale agent interactions.

\paragraph{Intra-group Actor-Critic.}

At the intra-group level, 
the government and households act as the leader and followers, respectively. We capture the government-household interaction with explicit modeling and process competition among households through implicit modeling.
For the household agents, their policy is conditioned on the household private observation and the government's decision, which captures the follower's response under the leader's strategy, denoted as$\pi_\text{hh}^{j,n} = \pi_{\theta_\text{hh}^n}(a_\text{hh}^{j,n}|o_{\text{hh}}^{j,n},a_{\text{gov},t})$.
%
The government agent's policy is solely conditioned on its local observation, denoted by $\pi_{\text{gov}}^n = \pi_{\theta_{\text{gov}}^n}(a_{\text{gov}}^n|o_{\text{gov}}^n)$.

The household critic estimates the expected return based on its individual state-action pair, given by $Q_{\text{hh}}^n := Q_{\phi_\text{hh}^n}(o_{\text{hh}}^n,\bar a_{\text{hh}})$.
Considering the large number of household agents, 
it is impossible to incorporate all household actions into the government critic network. 
Therefore, we employ a mean-field representation to aggregate household behaviors: $\bar a_{\text{hh}}^n = M(a_{\text{hh}}^n) = \frac{1}{N_{\text{gov}}} \sum_{j=1}^{N_{\text{gov}}} a_{\text{hh}}^{j,n}$, where $N_{\text{gov}}$ denotes the number of households within group $n$. Thus, the intra-group government critic is given by $Q_{\text{gov}}^{n,intra} := Q_{\phi_{\text{gov}}^{intra}}(o_{\text{gov}}^n,\bar a_\text{hh})$.


\paragraph{Inter-group Actor-Critic.}
Apart from the interactions within a group, the governments in different groups also compete with each other. Therefore, we introduce an additional inter-group critic $Q_{\text{gov}}^{n,inter} := Q_{\phi_{\text{gov}}^{inter}}(o_{\text{gov}},a_{\text{gov}})$ as a complementary estimate of cross-group competition.
%
The complete government value estimation should combine the two levels of estimation. Here, we design $Q_{\text{gov}}^n = Q_{\text{gov}}^{n,intra} + \lambda Q_{\text{gov}}^{n,inter}$ as the government's critic in the MHTG, where $\lambda$ balances the relative importance between internal and external interactions.

Overall, the MGPPO decomposes high-dimensional multi-agent optimization into structured subproblems while preserving scalability and the environment's hierarchical game structure.

 
\subsection{Training Procedure}

Introducing inter-group competition induces a nested game structure and tightly couples intra- and inter-group dynamics. In particular, the capital tax rate $\tau$ influences capital allocation across groups, which further affects the wage rate $w_t$ and household labor decisions. These household responses then alter capital accumulation and each government’s competitiveness in attracting mobile capital. Such feedback loops make the environment highly non-stationary and difficult to optimize with standard MARL methods. Conventional MARL approaches struggle to capture the structured interactions in the game, while bilevel MARL methods have difficulty simultaneously modeling intra- and inter-group dependencies.

We observe that the primary challenge arises from inter-group competition. Governments must not only maintain stable intra-group economic dynamics but also adapt strategically to competing governments. To address this, we propose a bilevel learning framework integrating two components: \textit{hierarchical sampling} and \textit{curriculum learning}, as illustrated in Figure~\ref{fig:pipeline}(b). Hierarchical sampling enforces the leader--follower structure by operating governments and households at different time scales. Curriculum learning gradually increases the strength of inter-group competition, allowing policies to stabilize before adapting to strategic interactions. Full implementation details are provided in Appendix~\ref{app:alg}.

\paragraph{Hierarchical Sampling} \label{sec:hs}
To preserve the leader--follower structure within each group, we introduce a \textit{hierarchical sampling} strategy that separates the decision time scales of governments and households. Households operate at a fine scale, sampling observations and acting at every time step, while the government updates its policy at a coarser interval of $n_{\text{gov}} > 1$ steps. During each government decision interval, the government policy remains fixed, allowing household behaviors to evolve under a stationary policy and reveal stable response patterns. Data are collected according to this schedule for each group. This design enables the government to better anticipate household responses and reinforces the intended leader--follower learning structure.

\paragraph{Curriculum Learning} \label{cl}
Curriculum learning mechanism gradually introduces inter-group competition into the intra-group leader--follower game. It
controls the intensity of inter-group competition through the capital mobility parameter $\phi$, defined as 
   $ \phi = \min\left(0.001 \times \text{episode},\ 1\right).$
At early training stages, $\phi$ is close to zero, implying limited capital mobility between groups and effectively decoupling groups, allowing the model to focus primarily on the intra-group leader--follower game. As training progresses and the intra-group strategies largely stabilize, $\phi$ gradually increases, strengthening inter-group competition and encouraging governments to adapt their policies in response to rivals. 
In implementation, we alternate between intra-group and inter-group updates. For each episode, we first suppress inter-group competition and train governments and households within each group under the leader--follower structure. Next, we fix the learned intra-group policies, set $\phi$ according to the curriculum, and update the inter-group policy. This alternating procedure stabilizes training while enabling gradual adaptation to competition.

\section{Experiments and Analysis}
\label{sec:exp}

We evaluate the proposed MGPPO framework through comparative experiments and a scalability analysis. The evaluation considers survival years, GDP, household consumption, and income and wealth inequality. One environment step corresponds to one year, and each episode is capped at 300 years. An episode may terminate earlier when the environment's capital, household-asset, inequality, or numerical-validity conditions are violated. The detail can be found in Appendix \ref{add:c1}. Survival years therefore measure how long a learned policy maintains a feasible economic trajectory within the evaluation horizon. GDP denotes average annual group output, while the Gini coefficients measure inequality among households within a group; lower Gini values indicate greater equality.
These results are calculated from 20 random evaluation seeds .  Values in parentheses denote the half-width of a two-sided 95\% Student-$t$ confidence interval across the 20 seed-level observations.  Additionally, within-episode action dynamics are presented in Appendix~\ref{sec:within_episode_actions}.
\subsection{Comparative Experiments}
\label{sec:exp.comparative}

We compare MGPPO with three on-policy multi-agent baselines: IPPO, MAPPO, and MAA2C. We examine the learned policies and the resulting economic outcomes for $N=2$ and $N=5$ groups.

\paragraph{Economic-indicator and reward analysis.}

Table~\ref{tab:economic-metrics-n2-n5} compares economic outcomes and survival years across algorithms.
For $N=2$, MGPPO reaches the 300-year limit, as do IPPO and MAA2C. Compared with IPPO, it increases mean GDP by 4.1\% (from $5.16\times10^6$ to $5.37\times10^6$) and reduces the mean income and wealth Gini coefficients by 5.0\% and 4.9\%, respectively. Its mean household consumption is 4.8\% lower. MGPPO records the lowest mean Gini coefficients and highest GDP among the four methods.
For $N=5$, MGPPO maintains a mean survival time of 300 years, compared with 297.2, 285.4, and 275.9 years for IPPO, MAPPO, and MAA2C, respectively. It also records the lowest income and wealth Gini coefficients and the highest mean household consumption. However, its mean GDP of $5.21\times10^6$ is the lowest of the four methods, whereas MAA2C attains the highest value of $5.64\times10^6$. These results reveal a trade-off between economic output, consumption, equality, and survival rather than uniform superiority across all indicators.

\begin{figure}[H]
\centering
\begin{subfigure}[t]{0.35\textwidth}
    \centering
    \includegraphics[width=\linewidth]{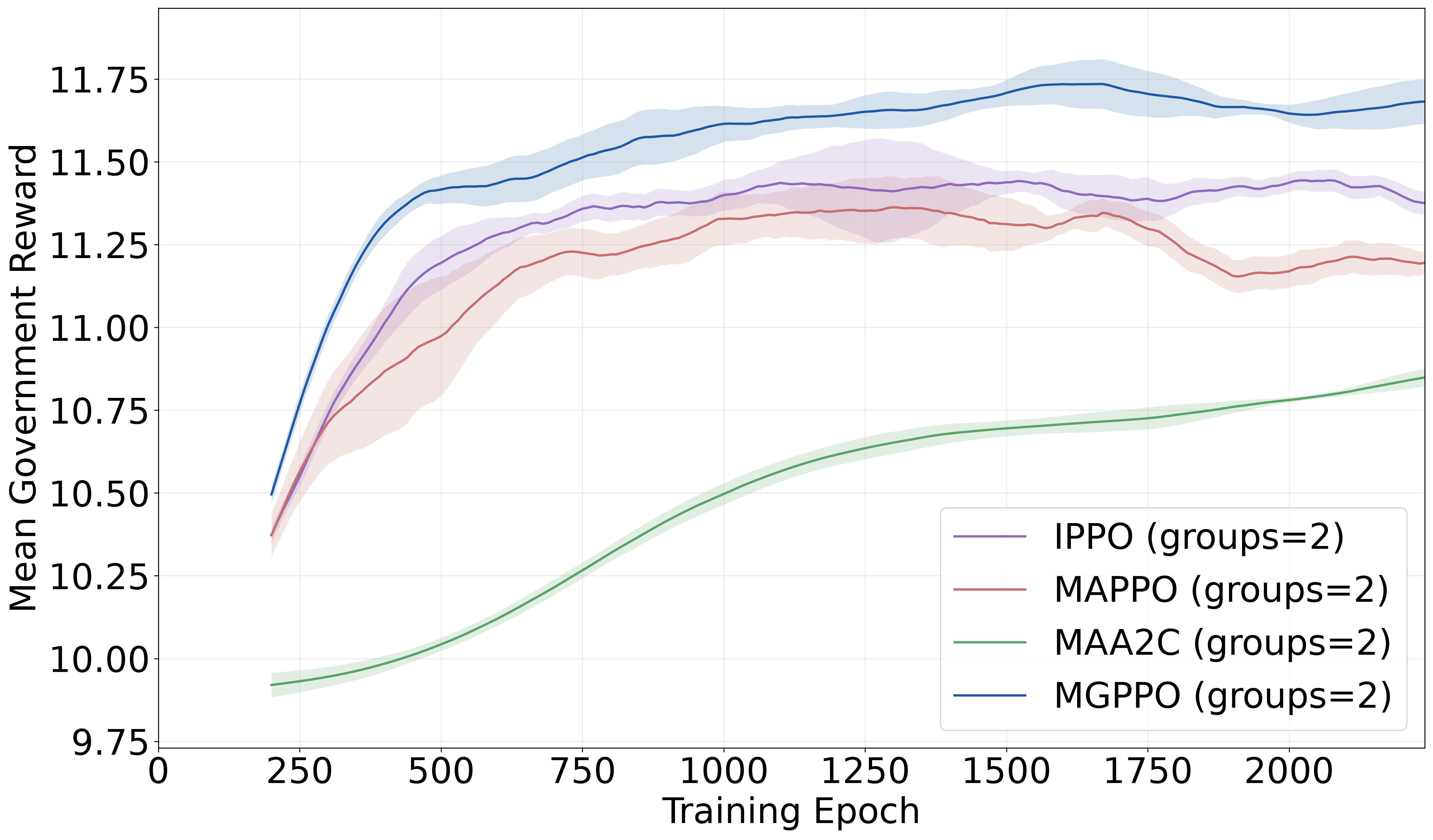}
    \caption{Government reward.}
    \label{fig:government-reward}
\end{subfigure} 
\begin{subfigure}[t]{0.35\textwidth}
    \centering
    \includegraphics[width=\linewidth]{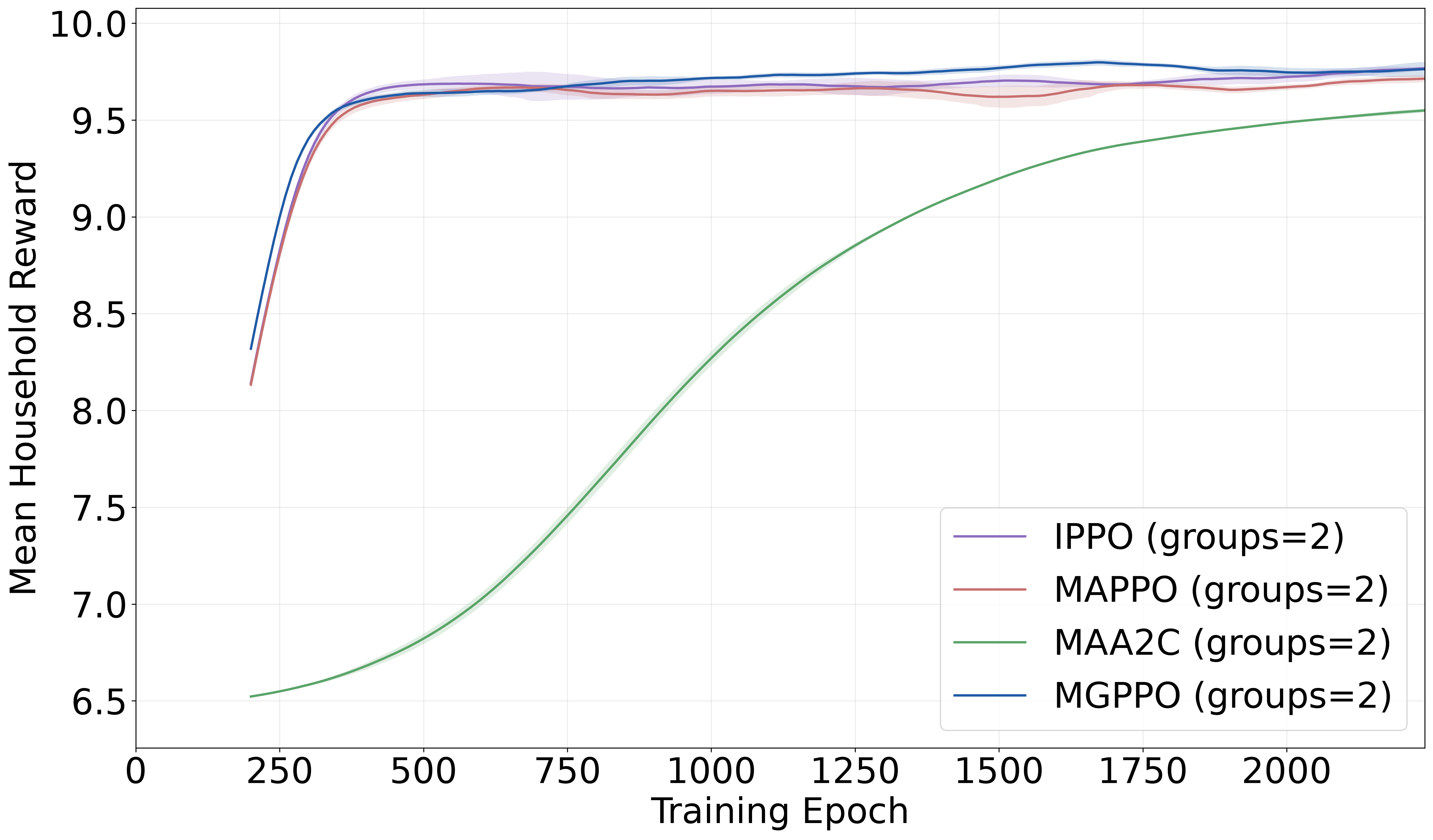}
    \caption{Household reward.}
    \label{fig:household-reward}
\end{subfigure}
\caption{Mean government and household rewards during training for $N=2$.}
\label{fig:reward-comparison}
\end{figure}

\begin{table}[H]
\centering
\scriptsize
\setlength{\tabcolsep}{3.2pt}
\renewcommand{\arraystretch}{1.08}
\resizebox{\textwidth}{!}{%
\begin{tabular}{@{}lcccccccc@{}}
\toprule
\textbf{Metric} &
\multicolumn{2}{c}{\textbf{IPPO}} &
\multicolumn{2}{c}{\textbf{MAPPO}} &
\multicolumn{2}{c}{\textbf{MAA2C}} &
\multicolumn{2}{c}{\textbf{MGPPO}} \\
\cmidrule(lr){2-3} \cmidrule(lr){4-5} \cmidrule(lr){6-7} \cmidrule(lr){8-9}
& \textbf{$N=2$} & \textbf{$N=5$}
& \textbf{$N=2$} & \textbf{$N=5$}
& \textbf{$N=2$} & \textbf{$N=5$}
& \textbf{$N=2$} & \textbf{$N=5$} \\
\midrule
Avg. consumption ($10^3$) $\uparrow$
& \textbf{26.02(0.60)} & 12.74(0.21)
& 9.35(0.21) & 17.49(0.70)
& 12.03(0.25) & 10.25(0.43)
& 24.78(0.49) & \textbf{20.16(0.24)} \\
Survival years $\uparrow$
& \textbf{300.0(0.0)} & 297.2(5.9)
& 297.2(5.8) & 285.4(18.7)
& \textbf{300.0(0.0)} & 275.9(24.3)
& \textbf{300.0(0.0)} & \textbf{300.0(0.0)} \\
Income Gini (\%) $\downarrow$
& 27.72(0.20) & 36.01(0.16)
& 36.07(0.20) & 36.53(0.42)
& 35.28(0.20) & 36.69(0.49)
& \textbf{26.34(0.16)} & \textbf{29.36(0.14)} \\
Wealth Gini (\%) $\downarrow$
& 22.37(1.35) & 26.90(0.66)
& 26.56(0.94) & 27.66(1.07)
& 26.28(0.99) & 27.79(0.84)
& \textbf{21.27(1.07)} & \textbf{21.95(0.79)} \\
Avg. GDP ($10^6$) $\uparrow$
& 5.16(0.13) & 5.29(0.09)
& 5.32(0.12) & 5.37(0.22)
& 5.35(0.12) & \textbf{5.64(0.25)}
& \textbf{5.37(0.12)} & 5.21(0.08) \\
\bottomrule
\end{tabular}%
}
\caption{Economic outcomes for $N=2$ and $N=5$ over 20 evaluation seeds.}
\label{tab:economic-metrics-n2-n5}
\end{table}
Figure~\ref{fig:reward-comparison} shows the reward trajectories for $N=2$. At the final recorded training epoch, MGPPO achieves the highest mean government reward, while its mean household reward remains close to that of IPPO. This pattern is consistent with MGPPO's leader--follower design, in which government policies influence household decisions and governments learn to account for those responses when optimizing their own rewards.

\paragraph{Action analysis.}

Table~\ref{tab:group-mean-actions-n2-n5} reports the normalized actions. For $N=2$, MGPPO selects higher income- and asset-tax parameter actions than IPPO but lower income-tax slope and spending-ratio actions. MGPPO records lower income and wealth Gini coefficients, a higher mean GDP ($5.37\times10^6$ versus $5.16\times10^6$), and lower mean consumption. Because total transfers are defined as $G_t=\eta_tY_t$, the spending ratio alone does not determine the absolute transfer amount. MGPPO also selects a higher saving-ratio action than IPPO (19.93\% versus 15.53\%), leaving a smaller share of current resources available for consumption.

For $N=5$, MGPPO combines the lowest income-tax parameter action with the highest asset-tax parameter, income-tax slope, spending-ratio, and capital-tax parameter actions. It records the highest mean consumption ($20.16\times10^3$), the lowest income and wealth Gini coefficients (29.36\% and 21.95\%), and a mean survival time of 300 years. Its high spending-ratio action and low saving-ratio action (24.82\%) allocate more available resources to current consumption.
Despite selecting the highest working-hour action (60.96\%), MGPPO records the lowest mean GDP ($5.21\times10^6$) at $N=5$. Thus, the highest working-hour action does not correspond to the highest GDP in this setting. Output depends on both effective labor and productive capital, while capital taxation affects capital availability and inter-group flows. 

\subsection{Scalability and Robustness Analysis}
\label{sec:exp.scalability}
\begin{figure}[t]
\centering
\begin{subfigure}[t]{0.24\textwidth}
    \centering
    \includegraphics[width=\linewidth]{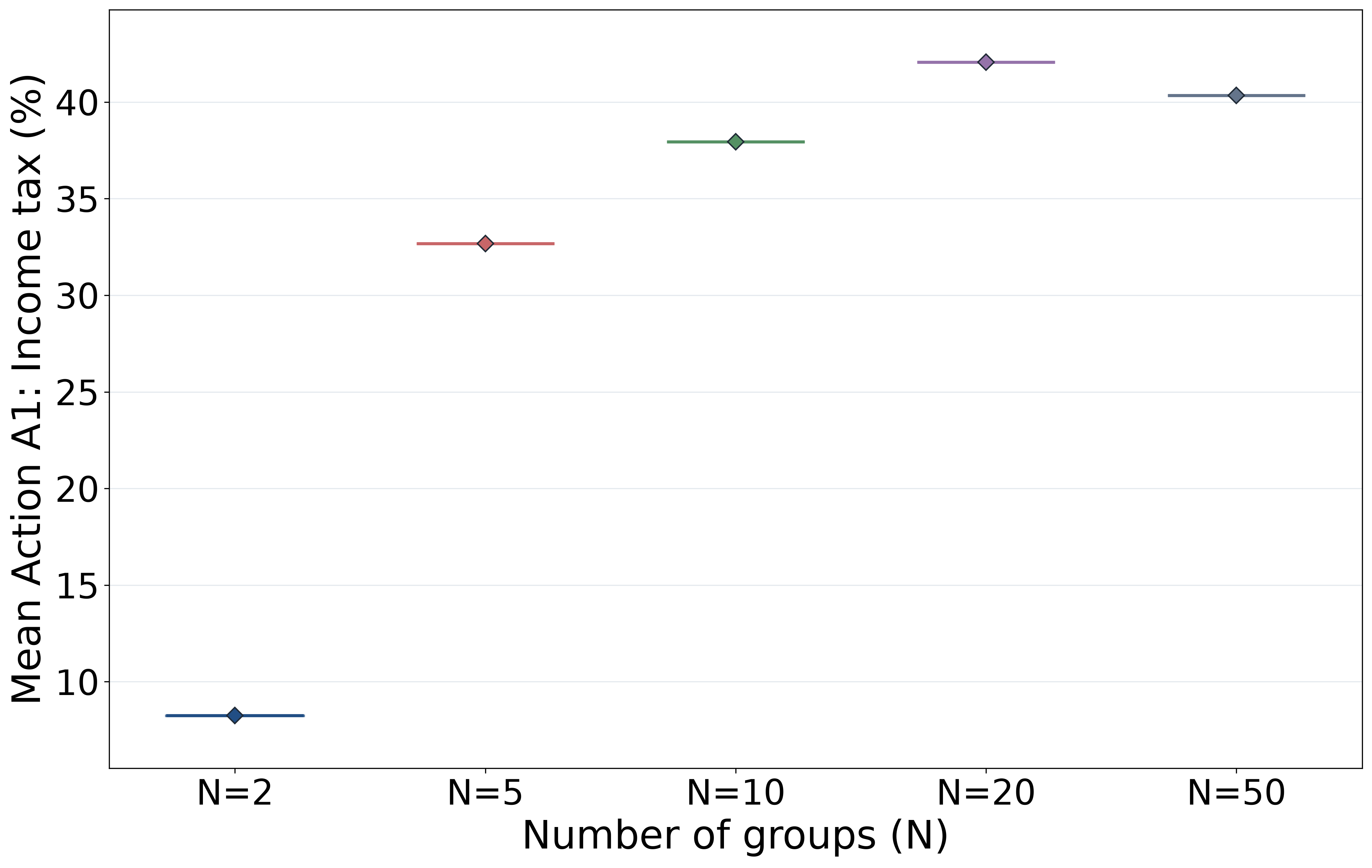}
    \caption{Income-tax parameter.}
    \label{fig:action-a1}
\end{subfigure}\hfill
\begin{subfigure}[t]{0.24\textwidth}
    \centering
    \includegraphics[width=\linewidth]{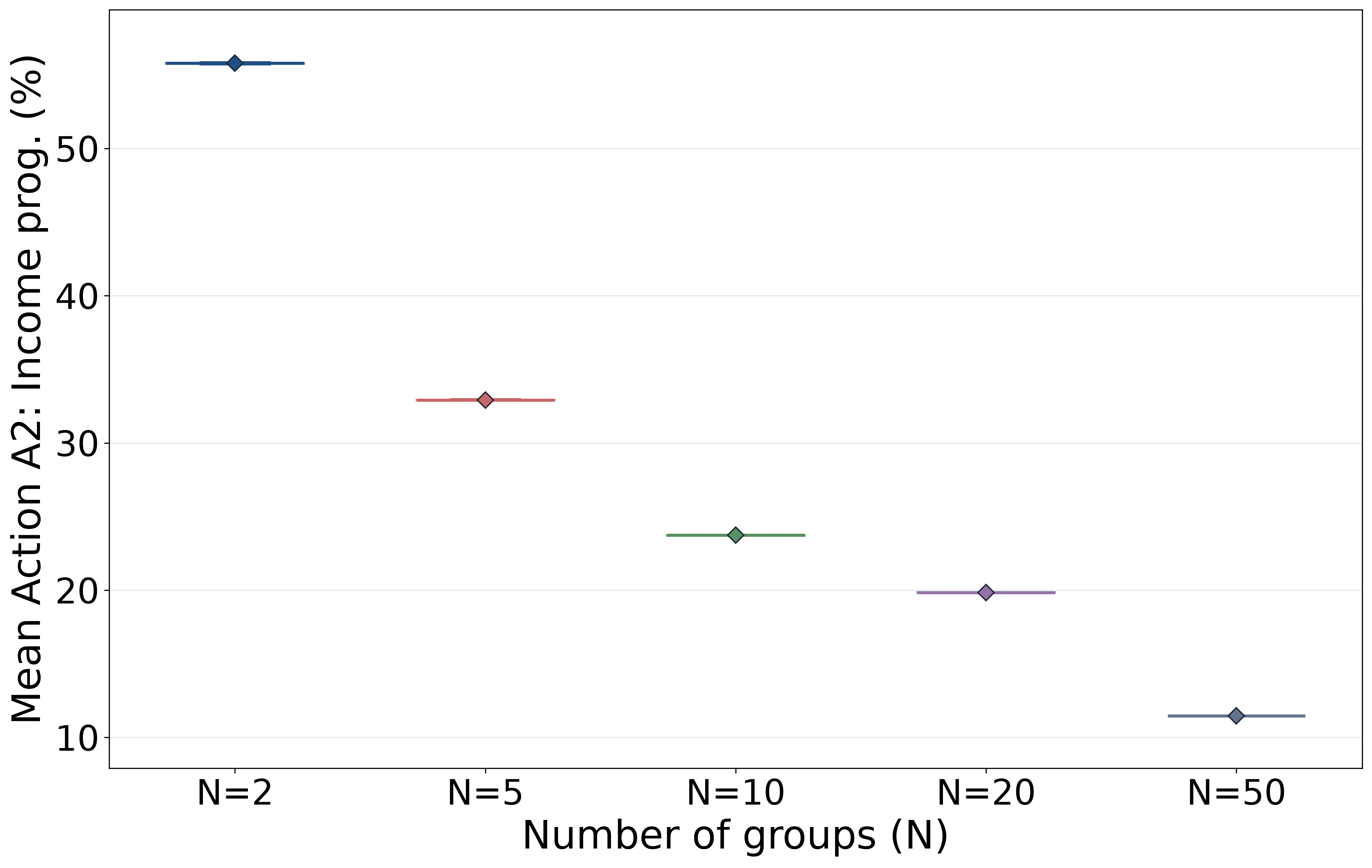}
    \caption{Income-tax slope.}
    \label{fig:action-a2}
\end{subfigure}\hfill
\begin{subfigure}[t]{0.24\textwidth}
    \centering
    \includegraphics[width=\linewidth]{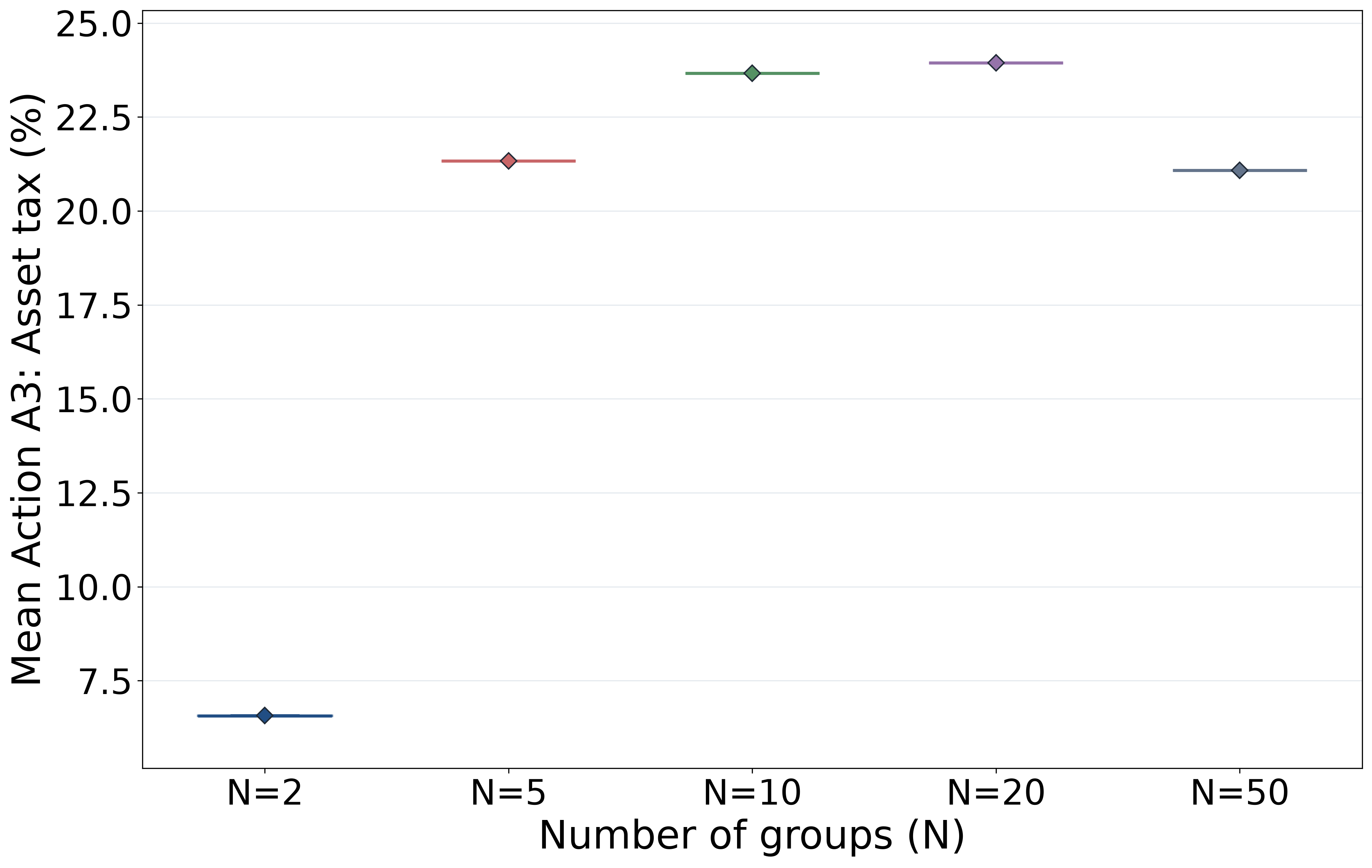}
    \caption{Asset-tax parameter.}
    \label{fig:action-a3}
\end{subfigure}\hfill
\begin{subfigure}[t]{0.24\textwidth}
    \centering
    \includegraphics[width=\linewidth]{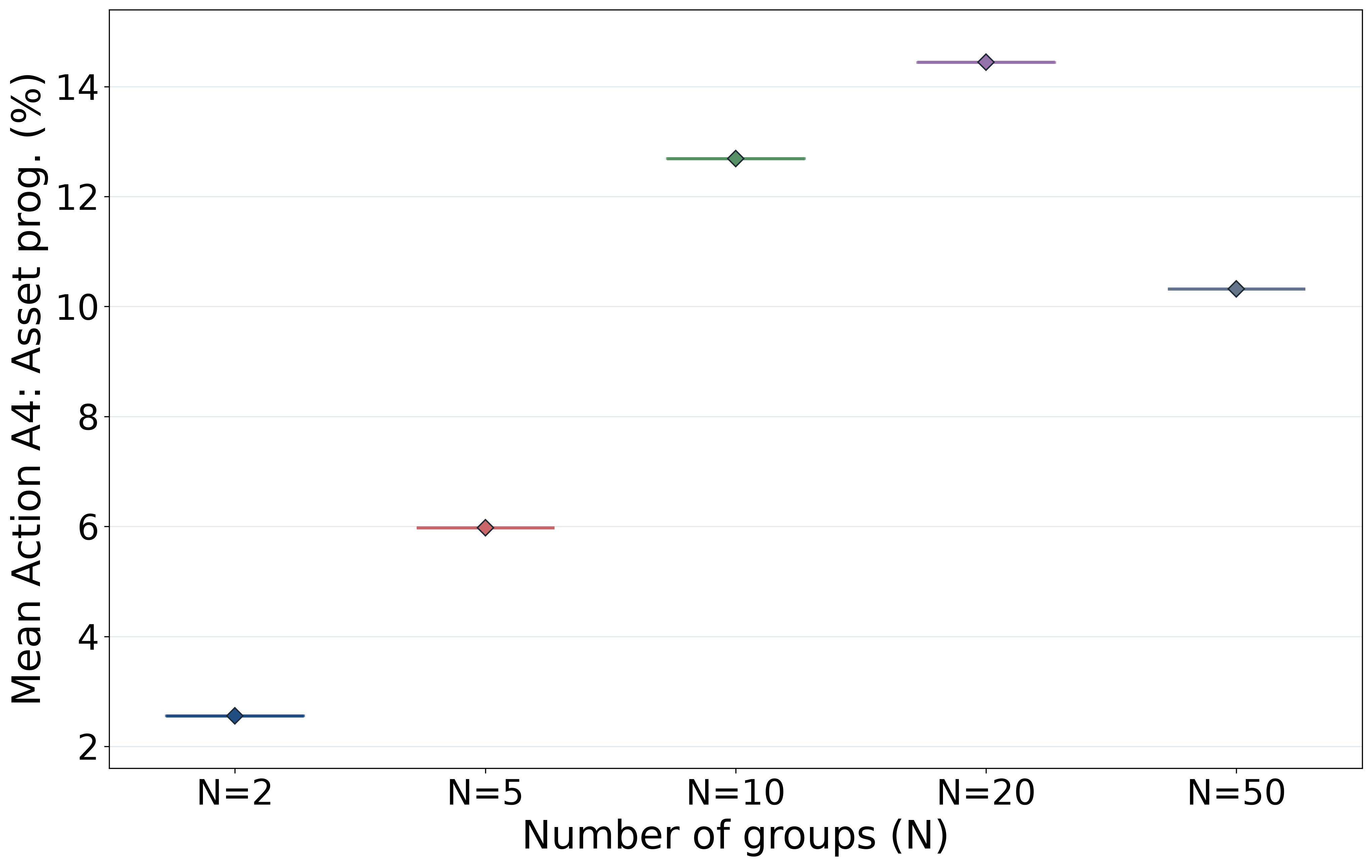}
    \caption{Asset-tax slope.}
    \label{fig:action-a4}
\end{subfigure}

\vspace{0.8em}

\begin{subfigure}[t]{0.24\textwidth}
    \centering
    \includegraphics[width=\linewidth]{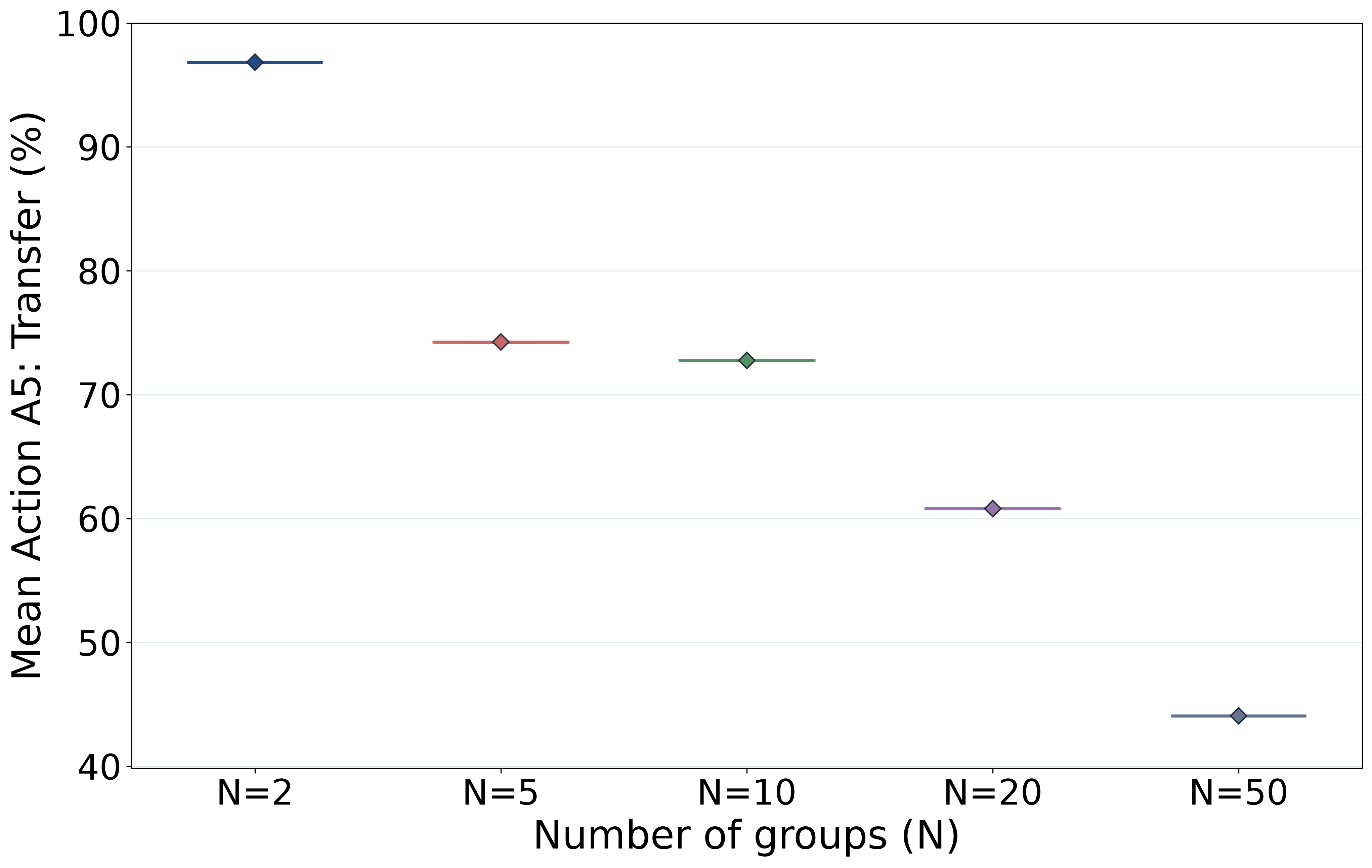}
    \caption{Spending ratio.}
    \label{fig:action-a5}
\end{subfigure}\hfill
\begin{subfigure}[t]{0.24\textwidth}
    \centering
    \includegraphics[width=\linewidth]{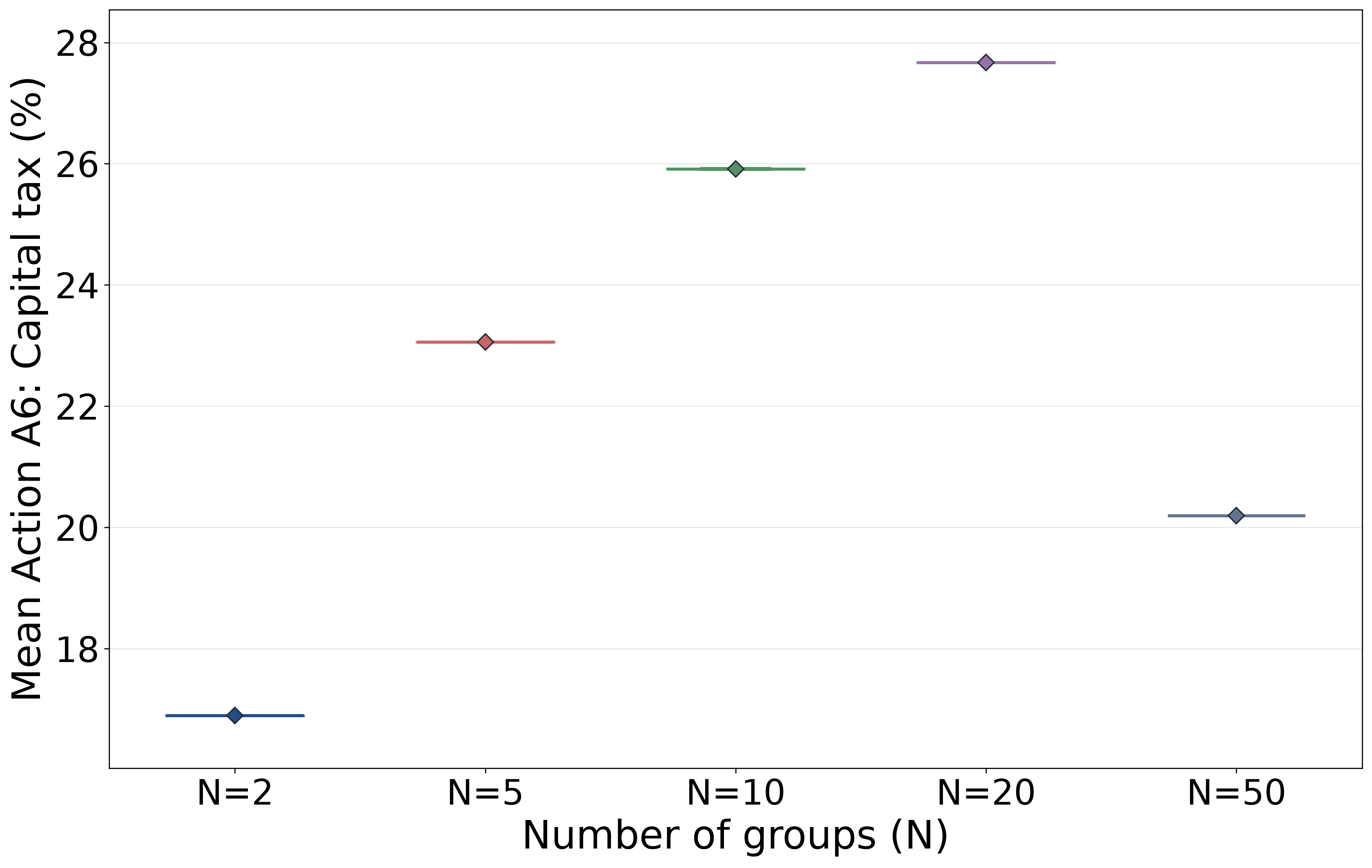}
    \caption{Capital-tax parameter.}
    \label{fig:action-a6}
\end{subfigure}\hfill
\begin{subfigure}[t]{0.24\textwidth}
    \centering
    \includegraphics[width=\linewidth]{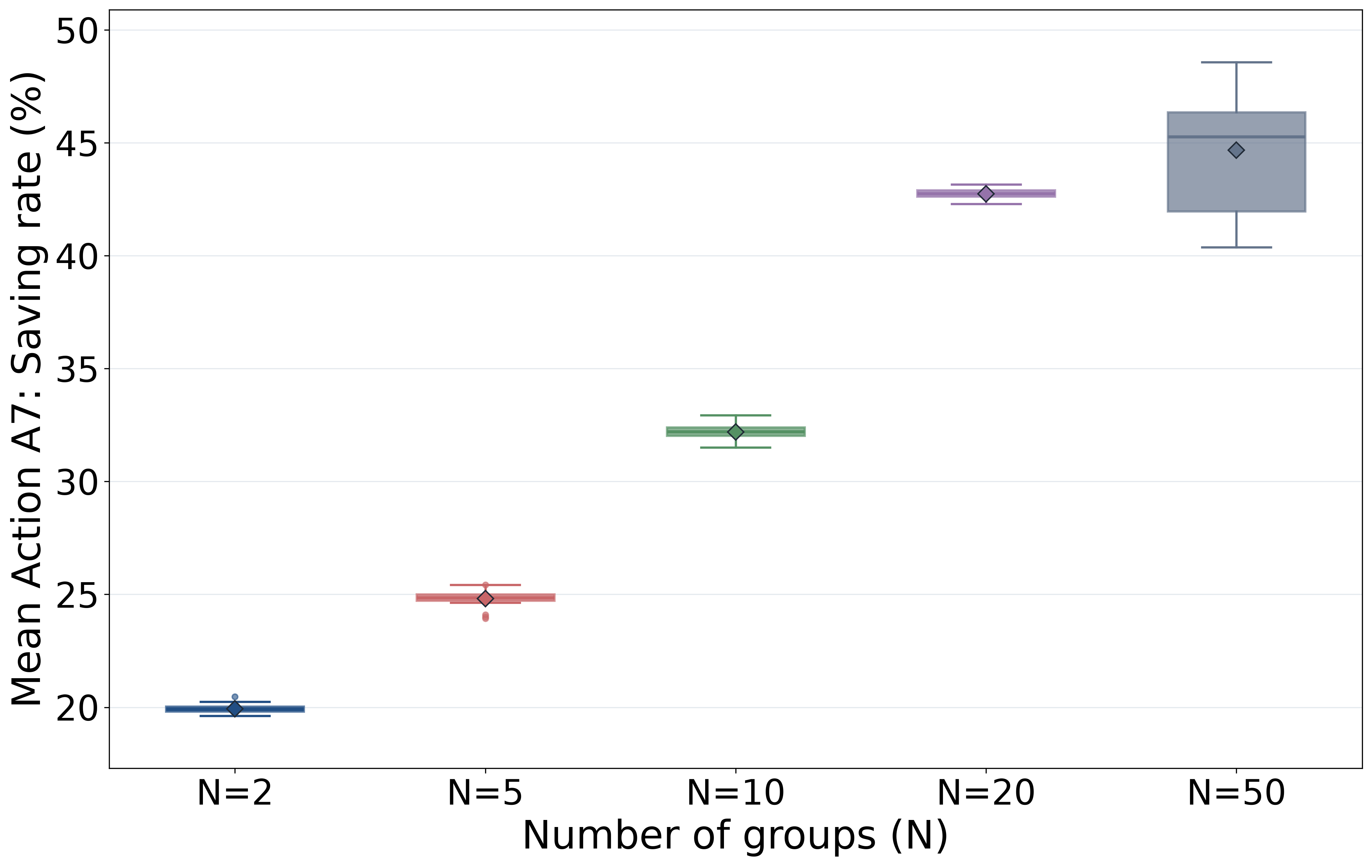}
    \caption{Saving ratio.}
    \label{fig:action-a7}
\end{subfigure}\hfill
\begin{subfigure}[t]{0.24\textwidth}
    \centering
    \includegraphics[width=\linewidth]{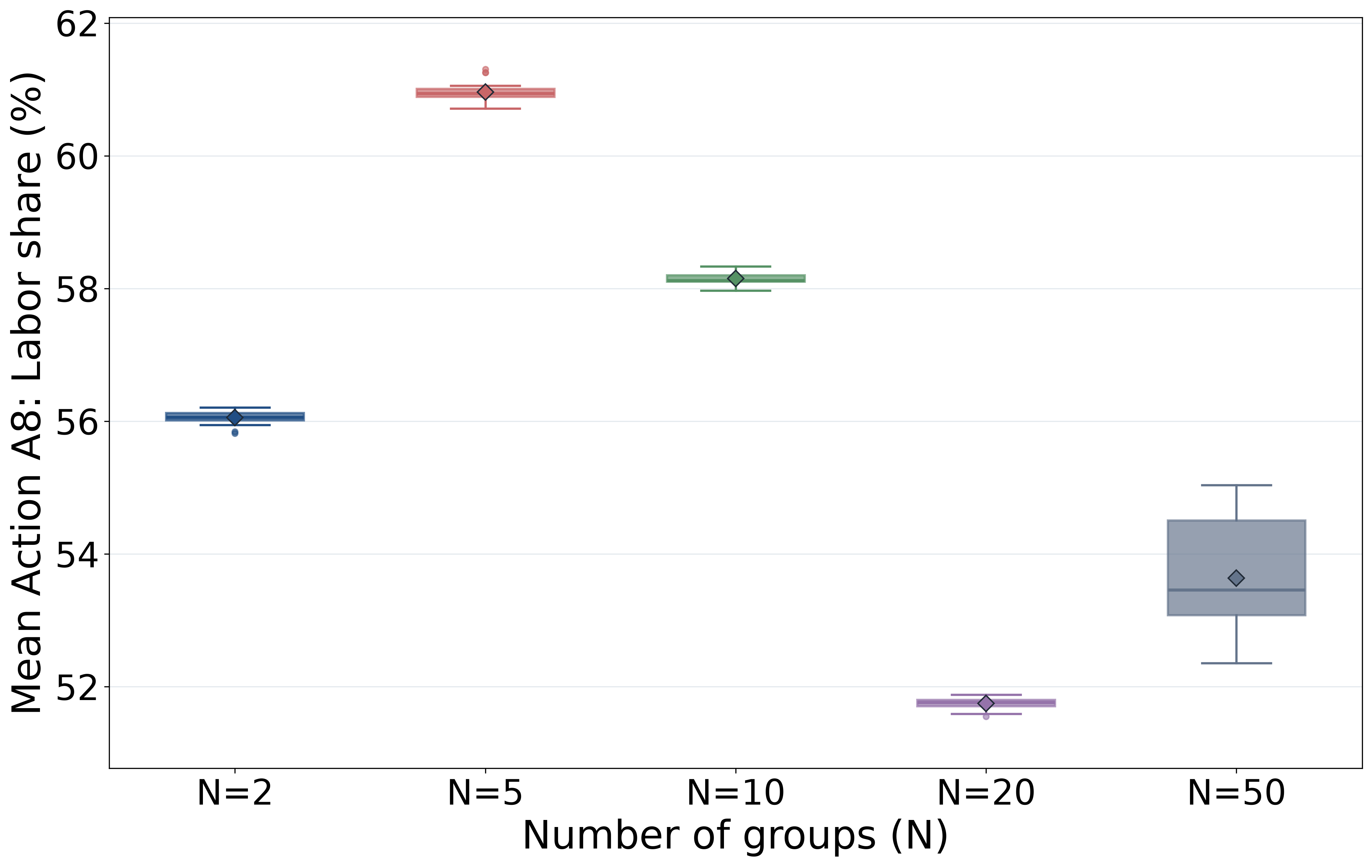}
    \caption{Working-hour action.}
    \label{fig:action-a8}
\end{subfigure}

\caption{MGPPO's government and household actions
across group counts $N\in\{2,5,10,20,50\}$.}
\label{fig:mgppo-action-boxplots}
\end{figure}

We examine MGPPO across different group sizes. Figure~\ref{fig:mgppo-action-boxplots} presents the distributions of eight government and household actions across  $N\in\{2,5,10,20,50\}$. As the number of groups increases across the smaller and intermediate settings, the income-tax parameter, asset-tax parameter, and capital-tax parameter actions generally shift upward, while the income-tax slope and spending ratio actions shift downward. This pattern reflects a change in the fiscal-policy mix rather than a uniform increase in taxation or spending. The upward shift in the capital-tax parameter action does not, by itself, indicate the direction of capital flows, which depends on each group's tax rate relative to the market average.
The household saving ratio action also shifts upward as the government spending ratio action declines. This pattern is consistent with a shift toward private saving as the spending ratio action falls. However, the spending ratio action does not measure the transfer received by each household, which also depends on output. Working-hour actions do not follow a monotonic trend, indicating that saving and labor decisions do not change in the same way as group size increases.
At the largest group size, the government action distributions shift downward but remain concentrated, while the household saving ratio action shows greater dispersion. Overall, MGPPO adjusts its policy mix across group sizes while maintaining concentrated government action distributions, demonstrating strong stability and scalability at the level of policy actions.

\begin{table}[t]
\centering
\scriptsize
\setlength{\tabcolsep}{3.2pt}
\renewcommand{\arraystretch}{1.08}
\resizebox{\textwidth}{!}{%
\begin{tabular}{@{}lcccccccc@{}}
\toprule
\textbf{Normalized action (\%)} & \multicolumn{2}{c}{\textbf{IPPO}} & \multicolumn{2}{c}{\textbf{MAPPO}} & \multicolumn{2}{c}{\textbf{MAA2C}} & \multicolumn{2}{c}{\textbf{MGPPO}} \\
\cmidrule(lr){2-3} \cmidrule(lr){4-5} \cmidrule(lr){6-7} \cmidrule(lr){8-9}
& \textbf{$N=2$} & \textbf{$N=5$} & \textbf{$N=2$} & \textbf{$N=5$} & \textbf{$N=2$} & \textbf{$N=5$} & \textbf{$N=2$} & \textbf{$N=5$} \\
\midrule
Income-tax parameter $\tau_i$ & 3.84(0.0135) & 39.48(0.0010) & 39.48(0.0024) & 41.27(0.0090) & 40.52(0.0005) & 40.24(0.0009) & 8.25(0.0046) & 32.67(0.0018) \\
Income-tax slope $\xi_i$ & 63.72(0.0832) & 11.36(0.0011) & 11.21(0.0040) & 10.76(0.0039) & 12.53(0.0032) & 10.41(0.0013) & 55.79(0.0180) & 32.92(0.0110) \\
Asset-tax parameter $\tau_a$ & 2.41(0.0065) & 19.12(0.0007) & 19.55(0.0021) & 19.68(0.0014) & 19.80(0.0004) & 20.15(0.0008) & 6.57(0.0024) & 21.33(0.0011) \\
Asset-tax slope $\xi_a$ & 4.60(0.0033) & 10.27(0.0004) & 9.87(0.0029) & 9.77(0.0011) & 10.34(0.0003) & 10.01(0.0001) & 2.56(0.0011) & 5.98(0.0009) \\
Spending ratio $\eta_t$ & 98.11(0.0064) & 47.94(0.0051) & 48.74(0.0242) & 44.41(0.0181) & 51.24(0.0116) & 41.78(0.0071) & 96.84(0.0028) & 74.25(0.0090) \\
Capital-tax parameter $\tau_{k,t}^{(n)}$ & 16.01(0.0035) & 20.52(0.0005) & 19.52(0.0010) & 21.61(0.0049) & 20.18(0.0003) & 19.96(0.0003) & 16.90(0.0015) & 23.06(0.0017) \\
\midrule
Saving ratio $p_t$ & 15.53(0.0218) & 55.62(0.0716) & 67.91(0.1048) & 38.97(0.1498) & 57.88(0.0392) & 65.31(0.0349) & 19.93(0.0948) & 24.82(0.1931) \\
Working-hour action $h_t$ & 59.04(0.0809) & 52.71(0.0330) & 49.95(0.0616) & 58.62(0.0673) & 53.65(0.0196) & 52.17(0.0206) & 56.06(0.0488) & 60.96(0.0753) \\
\bottomrule
\end{tabular}%
}
\caption{Group-averaged government and household actions for $N=2$ and $N=5$.}
\label{tab:group-mean-actions-n2-n5}
\end{table}

\section{Conclusion}

In this study, we developed a multi-group hierarchical tax game environment and proposed MGPPO to capture government–household interactions and competition among governments. The proposed method integrates \textit{hierarchical sampling} with \textit{curriculum learning} to optimize policies in this setting. In the two-group setting, MGPPO achieves a 4.1\% higher mean GDP than IPPO while reducing the income and wealth Gini coefficients by 5.0\% and 4.9\%, respectively. In the five-group setting, MGPPO reaches the 300-year evaluation horizon in all 20 runs and records the lowest inequality and highest household consumption among the compared methods. For $N\in\{2,5,10,20\}$, the scalability experiments demonstrate that MGPPO maintains stable policy performance across different group sizes, confirming its robustness and scalability. Overall, the results reveal a trade-off between distributional outcomes and aggregate output.
Future work will investigate the underlying fiscal mechanisms and strengthen economic robustness as the numbers of groups and households increase.

\bibliographystyle{plain}
\bibliography{example_paper}

@book{fudenberg1991game,
  title={Game theory},
  author={Fudenberg, Drew and Tirole, Jean},
  year={1991},
  publisher={MIT press}
}

@article{de2020independent,
  title={Is independent learning all you need in the starcraft multi-agent challenge?},
  author={De Witt, Christian Schroeder and Gupta, Tarun and Makoviichuk, Denys and Makoviychuk, Viktor and Torr, Philip HS and Sun, Mingfei and Whiteson, Shimon},
  journal={arXiv preprint arXiv:2011.09533},
  year={2020}
}

@article{yu2022surprising,
  title={The surprising effectiveness of ppo in cooperative multi-agent games},
  author={Yu, Chao and Velu, Akash and Vinitsky, Eugene and Gao, Jiaxuan and Wang, Yu and Bayen, Alexandre and Wu, Yi},
  journal={Advances in neural information processing systems},
  volume={35},
  pages={24611--24624},
  year={2022}
}

@article{zhong2024heterogeneous,
  title={Heterogeneous-agent reinforcement learning},
  author={Zhong, Yifan and Kuba, Jakub Grudzien and Feng, Xidong and Hu, Siyi and Ji, Jiaming and Yang, Yaodong},
  journal={Journal of Machine Learning Research},
  volume={25},
  number={32},
  pages={1--67},
  year={2024}
}

@book{von2010market,
  title={Market Structure and Equilibrium},
  author={Von Stackelberg, Heinrich},
  year={2011},
  publisher={Springer Science \& Business Media},
  note={First published in 1934, this is the translation.}
}

@article{colson2007overview,
  title={An overview of bilevel optimization},
  author={Colson, Beno{\^\i}t and Marcotte, Patrice and Savard, Gilles},
  journal={Annals of operations research},
  volume={153},
  number={1},
  pages={23--56},
  year={2007},
  publisher={Springer}
}

@article{mirrlees1971exploration,
  title={An exploration in the theory of optimum income taxation},
  author={Mirrlees, James A},
  journal={The review of economic studies},
  volume={38},
  number={2},
  pages={175--208},
  year={1971},
  publisher={Wiley-Blackwell}
}

@article{wilson1999theories,
  title={Theories of tax competition},
  author={Wilson, John Douglas},
  journal={National tax journal},
  volume={52},
  number={2},
  pages={269--304},
  year={1999},
  publisher={The University of Chicago Press}
}

@book{ikeda2002dynamics,
  title={Dynamics of the mixed economy: Toward a theory of interventionism},
  author={Ikeda, Sanford},
  year={2002},
  publisher={Routledge}
}

@book{johnson2014mixed,
  title={Mixed economies welfare},
  author={Johnson, Norman},
  year={2014},
  publisher={Routledge}
}

@book{axtell2000agents,
  title={Why agents?: on the varied motivations for agent computing in the social sciences},
  author={Axtell, Robert},
  volume={17},
  year={2000},
  publisher={Center on Social and Economic Dynamics Washington, DC}
}

@article{zhang2021multi,
  title={Multi-agent reinforcement learning: A selective overview of theories and algorithms},
  author={Zhang, Kaiqing and Yang, Zhuoran and Ba{\c{s}}ar, Tamer},
  journal={Handbook of reinforcement learning and control},
  pages={321--384},
  year={2021},
  publisher={Springer}
}

@article{zheng2022ai,
  title={The AI Economist: Taxation policy design via two-level deep multiagent reinforcement learning},
  author={Zheng, Stephan and Trott, Alexander and Srinivasa, Sunil and Parkes, David C and Socher, Richard},
  journal={Science advances},
  volume={8},
  number={18},
  pages={eabk2607},
  year={2022},
  publisher={American Association for the Advancement of Science}
}

@article{mi2023taxai,
  title={Taxai: A dynamic economic simulator and benchmark for multi-agent reinforcement learning},
  author={Mi, Qirui and Xia, Siyu and Song, Yan and Zhang, Haifeng and Zhu, Shenghao and Wang, Jun},
  journal={arXiv preprint arXiv:2309.16307},
  year={2023}
}

@article{KEEN199733,
title = {Fiscal competition and the pattern of public spending},
journal = {Journal of Public Economics},
volume = {66},
number = {1},
pages = {33-53},
year = {1997},
issn = {0047-2727},
doi = {https://doi.org/10.1016/S0047-2727(97)00035-2},
url = {https://www.sciencedirect.com/science/article/pii/S0047272797000352},
author = {Michael Keen and Maurice Marchand}
}

@book{atkinson1980lectures,
  title={Lectures on public economics: Updated edition},
  author={Atkinson, Anthony B and Stiglitz, Joseph E},
  year={2015},
  publisher={Princeton University Press}
}

@article{ZODROW1986356,
title = {Pigou, Tiebout, property taxation, and the underprovision of local public goods},
journal = {Journal of Urban Economics},
volume = {19},
number = {3},
pages = {356-370},
year = {1986},
issn = {0094-1190},
doi = {https://doi.org/10.1016/0094-1190(86)90048-3},
url = {https://www.sciencedirect.com/science/article/pii/0094119086900483},
author = {George R. Zodrow and Peter Mieszkowski}
}

@article{10.2307/2118417,
    author = {Aiyagari, S. Rao},
    title = {Uninsured Idiosyncratic Risk and Aggregate Saving*},
    journal = {The Quarterly Journal of Economics},
    volume = {109},
    number = {3},
    pages = {659-684},
    year = {1994},
    month = {08},
    issn = {0033-5533},
    doi = {10.2307/2118417},
    url = {https://doi.org/10.2307/2118417},
    eprint = {https://academic.oup.com/qje/article-pdf/109/3/659/5203006/109-3-659.pdf},
}

@article{doi:10.1086/601445,
author = {Aiyagari, S. Rao},
title = {Optimal Capital Income Taxation with Incomplete Markets, Borrowing Constraints, and Constant Discounting},
journal = {Journal of Political Economy},
volume = {103},
number = {6},
pages = {1158-1175},
year = {1995},
doi = {10.1086/601445},

URL = { 
    
        https://doi.org/10.1086/601445
    
    

},
eprint = { 
    
        https://doi.org/10.1086/601445
    
    

}
}

@article{chamley1986optimal,
  title={Optimal taxation of capital income in general equilibrium with infinite lives},
  author={Chamley, Christophe},
  journal={Econometrica: Journal of the Econometric Society},
  pages={607--622},
  year={1986},
  publisher={JSTOR}
}

@article{benabou2002tax,
  title={Tax and education policy in a heterogeneous-agent economy: What levels of redistribution maximize growth and efficiency?},
  author={Benabou, Roland},
  journal={Econometrica},
  volume={70},
  number={2},
  pages={481--517},
  year={2002},
  publisher={Wiley Online Library}
}

@article{heathcote2017optimal,
  title={Optimal tax progressivity: An analytical framework},
  author={Heathcote, Jonathan and Storesletten, Kjetil and Violante, Giovanni L},
  journal={The Quarterly Journal of Economics},
  volume={132},
  number={4},
  pages={1693--1754},
  year={2017},
  publisher={Oxford University Press}
}

@article{curry2022analyzing,
  title={Analyzing micro-founded general equilibrium models with many agents using deep reinforcement learning},
  author={Curry, Michael and Trott, Alexander and Phade, Soham and Bai, Yu and Zheng, Stephan},
  journal={arXiv preprint arXiv:2201.01163},
  year={2022}
}

@article{trott2021building,
  title={Building a foundation for data-driven, interpretable, and robust policy design using the ai economist},
  author={Trott, Alexander and Srinivasa, Sunil and van der Wal, Douwe and Haneuse, Sebastien and Zheng, Stephan},
  journal={arXiv preprint arXiv:2108.02904},
  year={2021}
}

@phdthesis{shi2023deep,
  title={Deep reinforcement learning and macroeconomic modelling},
  author={Shi, Rui Aruhan},
  year={2023},
  school={University of Warwick}
}

@article{hinterlang2021optimal,
  title={Optimal monetary policy using reinforcement learning},
  author={Hinterlang, Natascha and T{\"a}nzer, Alina},
  year={2021},
  publisher={Deutsche Bundesbank Discussion Paper}
}

@article{chen2021deep,
  title={Deep reinforcement learning in a monetary model},
  author={Chen, Mingli and Joseph, Andreas and Kumhof, Michael and Pan, Xinlei and Zhou, Xuan},
  journal={arXiv preprint arXiv:2104.09368},
  year={2021}
}

@inproceedings{foerster2017stabilising,
  title={Stabilising experience replay for deep multi-agent reinforcement learning},
  author={Foerster, Jakob and Nardelli, Nantas and Farquhar, Gregory and Afouras, Triantafyllos and Torr, Philip HS and Kohli, Pushmeet and Whiteson, Shimon},
  booktitle={International conference on machine learning},
  pages={1146--1155},
  year={2017},
  organization={PMLR}
}

@inproceedings{zhang2020bi,
  title={Bi-level actor-critic for multi-agent coordination},
  author={Zhang, Haifeng and Chen, Weizhe and Huang, Zeren and Li, Minne and Yang, Yaodong and Zhang, Weinan and Wang, Jun},
  booktitle={Proceedings of the AAAI conference on artificial intelligence},
  volume={34},
  number={05},
  pages={7325--7332},
  year={2020}
}

\newpage
\appendix

\section{Bewley--Aiyagari Model} \label{app:ba-model}

The Bewley--Aiyagari model describes a minimal economic circulation system that includes four entities: household, government, firm, and financial intermediary.

\paragraph{Households}
In the economic system, a household get paid by working and spend money by consumption, forming a working--spending dynamics that powers economic circulation.
We define $i_t$, $a_t$, and $c_t$ as the income, wealth, and consumption of a household at time $t$, which represents the financial state of a household. We also denote $h_t$ and $w_t$ as the working hour and the wage rate, respectively. To capture heterogeneous characters of different households, we introduce productivity level $e_t$, which denotes the household's working efficiency. We assume $e_t$ follows an AR(1) process, given by
\begin{equation}
	\label{eq:et}
	\log e_{t+1} = \rho_e \log e_t + \sigma_e u_t,
\end{equation}
where $u_t \sim \mathcal{N}(0,1)$ denotes a standard normal shock, and $\rho_e \geq 0, \sigma_e \geq 0$ are the persistence and the volatility of the shock. Let $r_\text{interest}$ be the interest rate. A household's income in time $t$ can be given by
\begin{equation}
	\label{eq:it}  
	i_{t} = w_t h_t e_t + r_{\text{interest}} a_t,
\end{equation}
We consider three types of taxation policies in the model, which are income, asset, and consumption taxes, denoted by $T_{\text{inc}}(i_t)$, $T_{\text{ast}}(a_t)$, and $T_{\text{cons}}(c_t)$, respectively. Additionally, let \(G_t\) denote total government transfers and \(N\) the number of households receiving those transfers. Note that these tax policies are adjustable functions of a household's financial status. Based on the income, consumption, and tax, we can derive the wealth accumulation relation as
\begin{equation}
	\label{eq:at}
	a_{t+1} = i_t - T_{\text{inc}}(i_t) + a_t - T_{\text{ast}}(a_t) - c_t - T_{\text{cons}}(c_t)  +\frac{G_t}{N} .
\end{equation}
Let $p_t$ denote the household's saving ratio to simplify the complex household's working--spending dynamics in \eqref{eq:et}-\eqref{eq:at} and training, which is given by
\begin{equation}
	\label{eq:pt}
	p_t = \frac{a_{t+1}}{i_t - T_{\text{inc}}(i_t) + \sout{a_t - T_{\text{ast}}(a_t)}  + \frac{G_t}{N}}.
\end{equation}
This simplifies the wealth evolution \eqref{eq:at} to 
\begin{equation}
	\label{eq:at1}
	a_{t+1} = a_t - T_{\text{ast}}(a_t) + p_t \left( i_t - T_{\text{inc}}(i_t) +  \frac{G_t}{N} \right).
\end{equation}

\paragraph{Government} 
A government is responsible for taxing, social spending, and capital regulation. 
For taxation, we consider three types of tax policies: consumption, income, and asset taxes. In practice, the consumption tax can be modeled with a fixed rate $\tau_s$ \cite{chamley1986optimal}, i.e.,
\begin{equation}
\label{eq:T_cons}
    T_{\text{cons}}(c_t) = \tau_s c_t.
\end{equation}
The income tax $T_{\text{inc}}(i_t)$ and the asset tax $T_{\text{ast}}(a_t)$ are widely adopted by the nonlinear HSV \cite{benabou2002tax,heathcote2017optimal}, given by
\begin{equation}
\label{eq:T_inc}
    T_{\text{inc}}(i_t) = i_t - \frac{1-\tau_i}{1-\xi_i} (i_t)^{1-\xi_i},
\end{equation}
\begin{equation}
\label{eq:T_ast}
    T_{\text{ast}}(a_t) = a_t - \frac{1-\tau_a}{1-\xi_a} (a_t)^{1-\xi_a}.
\end{equation}
Here, $\tau_i$ and $\xi_i$ control the average level and the slope of the marginal income tax; similarly, $\tau_a$ and $\xi_a$ determine the average level and the slope of the asset tax. 

In its role of social spending, the government may issue debt into the economic system and must service this debt by paying interest at a rate $r_{\text{interest}}$. Let $G_t$ and $B_t$ denote the government's social spending and debt at time $t$, respectively, which together characterize its fiscal state. 
Assuming there are $M$ households paying taxes. With a slight abuse of notation, we use superscript to index households and denote $i^j_{t}$, $a^j_{t}$, $c^j_{t}$ as the $j$-th household's income, wealth, and consumption at time $t$. The government’s debt evolves according to
\begin{equation}
	\label{eq:Bt}
		B_{t+1} = ( 1+r_{\mathrm{interest}} ) B_t + G_t - \sum_{j=1}^M \left( T_{\mathrm{inc}}(i^j_{t})  + T_{\mathrm{ast}}(a^j_{t}) + T_{\mathrm{cons}}(c^j_{t}) \right).
\end{equation}
When introducing the spending ratio $\eta_t = \frac{G_t}{Y_t}$, the social spending $G_t$ can be represented by $\eta_t Y_t = \eta_t K_t^\alpha L_t^{1-\alpha}$ using Cobb-Douglas production function \eqref{eq:Yt}, which can be used to further simplify the debt dynamics \eqref{eq:Bt}.

Capital regulation arises in multi-group settings, where governments compete for mobile capital to enhance their economic welfare. This inter-group strategic behavior is captured through tax competition dynamics and is formally modeled in Appendix~\ref{app:zm-dynamics}.

\paragraph{Firm}
A firm abstracts all production units in the economic circulation. It hires households to conduct production and pays wages to households. We use the widely adopted Cobb-Douglas production function to summarize the production process:
\begin{equation}
	\label{eq:Yt}
	Y_t = K_t^\alpha L_t^{1-\alpha}
\end{equation}
where $K_t$ and $L_t$ are capital and labors used for production and $\alpha \in (0,1)$ is the capital elasticity. Suppose there are $M$ households participating the production, we have $L_t = \sum_{j=1}^M h^j_{t} e^j_{t}$ , representing the total effective labor supply. 
Using the assumption that the firm takes the marginal income from labor as the household’s wage rate $w_t$ (i.e., the market-clearing condition), we obtain
\begin{equation}
\label{eq:wt}
    w_t = \frac{\partial Y_t}{\partial L_t} 
    = (1-\alpha) \left( \frac{K_t}{L_t} \right)^\alpha 
    = (1-\alpha) \frac{K_t^\alpha}{\left( \sum_{j=1}^M e^j_{t} h^j_{t} \right)^\alpha}.
\end{equation}
It shows that all households face the same wage rate, which is endogenously determined by their collective labor supply and the available capital stock $K_t$.

\paragraph{Financial intermediary}

A financial intermediary abstracts institutions, such as banks, that collect households’ savings and allocate these funds to productive capital and government bonds. This allocation determines the market’s available capital for production in the subsequent period.
Assuming $M$ households deposit their savings into the intermediary, the evolution of aggregate capital and government debt satisfies
\begin{equation}
	\label{eq:Kt}
		K_{t+1} + B_{t+1} - \sum_{j=1}^M a^j_{t+1} = (1 + r_{\text{interest}})  K_t + (1 + r_{\text{interest}}) \left( B_t - \sum_{j=1}^M a^j_{t} \right).
\end{equation}
Here, $K_t$ denotes the capital stock held by the intermediary at time $t$, which corresponds to the total capital available to firms for production in \eqref{eq:Yt}. $B_t$ represents the government debt, and $a^j_{t}$ denotes the asset of household $j$ at time $t$. 
This equation balances the intermediary’s portfolio across periods and accounts for returns on capital and interest payments on government debt.

\section{Zodrow--Mieszkowski Dynamics} \label{app:zm-dynamics}

In a multi-group setting with $N$ governments, capital is treated as an imperfectly mobile factor of production. Investors observe the effective capital tax rates across all regions and reallocate capital to maximize after-tax returns. This process implies that capital tends to migrate from high-tax jurisdictions to low-tax ones, a phenomenon often referred to as the ``Race to the Bottom''.

We use superscript to denote the group index and let $K^{(n)}_{t}$ denote the capital stock in group $n$ at time $t$. The dynamics of capital flow are governed by the deviation of the local capital tax rate $\tau^{(n)}_{t}$ from the average tax rate of all groups $\bar{\tau}_t$. The law of motion for capital, adjusted for these flows, is formulated as:
\begin{equation}
\label{eq:zm_flow}
\Delta K_{t}^{(n),\mathrm{flow}}
=
-\phi_{e}
\left(
\tau_{k,t}^{(n)}-\bar{\tau}_{k,t}
\right)
K_{t}^{(n)},
\end{equation}
 
where $\bar{\tau}_{k,t}
=
\frac{
\sum_{m=1}^{N}\tau_{k,t}^{(m)}K_{t}^{(m)}
}{
\sum_{m=1}^{N}K_{t}^{(m)}
}.$ represents the market average tax rate. It is a classic inter-jurisdictional capital reallocation from the Zodrow--Mieszkowski tax competition model, widely adopted in the fiscal competition literature. Here, $\bar{\tau}_t = \frac{1}{N} \sum_{n=1}^N \tau^{(n)}_{t}$ represents the market average tax rate. It captures the standard ``Race to the Bottom'' phenomenon: capital flows out of high-tax jurisdictions and into low-tax ones proportionally to the tax differential $(\tau^{(n)}_t - \bar{\tau}_t)$ and the current capital stock $K^{(n)}_t$. The parameter $\phi > 0$ controls the intensity of capital mobility, and is treated as a fixed constant in standard models.

Incorporating this dynamic into the standard accumulation process, the aggregate capital for jurisdiction $i$ at the next time step is updated by:
\begin{equation}
	K^{(n),\text{real}}_{t+1} = K^{(n)}_{t+1} + \Delta K^{(n),\text{flow}}_{t},
\end{equation}
where $	K^{(n),\text{real}}_{t+1}$ represents the real available capital accumulation. This mechanism forces governments to strategically lower tax rates to attract or retain the tax base, often at the expense of public good provision or fiscal balance.

\section{Simulation Settings} \label{app:hyper}

\subsection{Termination Conditions}
\label{add:c1}

An episode ends when any of the following conditions is met:
\begin{itemize}
    \item The simulation reaches its 300-step limit. 
    \item Any group's next-period capital is non-positive.
    \item Any household's next-period assets become negative.
    \item A non-finite reward occurs.
    \item From step 21 onward, any group's wealth Gini coefficient exceeds 0.85 or its income Gini coefficient exceeds 0.469.
\end{itemize}
The step limit is treated as a time-limit termination, rather than an economic failure.

\subsection{Hyperparameters}

Tables~\ref{tab:hyperparameters1} and \ref{tab:hyperparameters2} summarize the default simulation settings and the training parameters of the four implemented algorithms. The number of groups and households can vary across experiments.

\begin{table}[htbp]
\centering
\caption{Default simulation and shared training settings.}
\label{tab:hyperparameters1}
\small
\renewcommand{\arraystretch}{1.15}
\begin{tabular}{lclc}
\toprule
Hyperparameter & Value & Hyperparameter & Value \\
\midrule
Number of groups & 2 & Households per group & 100 \\
Episode limit (steps) & 300 & Years per step & 1 \\
Government action interval & 10 & Inequality checks begin & Step 21 \\
Wealth Gini threshold & 0.85 & Income Gini threshold & 0.469 \\
Episodes per update & 5 & Evaluation episodes & 20 \\
Evaluation interval (updates) & 10 & Hidden units per layer & 256 \\
Activation & Tanh & Optimizer & Adam \\
Actor learning rate & $3\times10^{-6}$ & Critic learning rate & $3\times10^{-6}$ \\
Government discount factor & 0.995 & Household discount factor & 0.995 \\
Government GAE parameter & 0.95 & Household GAE parameter & 0.99 \\
Entropy coefficient & 0.01 & Maximum gradient norm & 0.5 \\
\bottomrule
\end{tabular}
\end{table}

\begin{table}[htbp]
\centering
\caption{Algorithm-specific training parameters.}
\label{tab:hyperparameters2}
\small
\renewcommand{\arraystretch}{1.15}
\begin{tabular}{lcccc}
\toprule
Hyperparameter & IPPO & MAPPO & MAA2C & MGPPO \\
\midrule
Update method & PPO & PPO & A2C & PPO \\
Policy-update passes & 10 & 10 & 1 & 10 \\
Minibatch size (steps) & 64 & 64 & Full batch & 64 \\
PPO clipping parameter & 0.2 & 0.2 & -- & 0.2 \\
KL threshold & 0.01 & 0.01 & -- & 0.01 \\
Inter-group critic weight & -- & -- & -- & 1.0 \\
\bottomrule
\end{tabular}
\end{table}

\subsection{Simulation Variables}
We summarize key variables in the economic simulation in Table~\ref{tab:var}.
\begin{table}[htbp] 
	\centering
    \caption{Key variables in the economic simulation environment}
	\renewcommand{\arraystretch}{1.1}
	\scalebox{0.85}{
	\begin{tabular}{llll}
		\toprule
		Variable & Meaning & Variable & Meaning \\
		\midrule
		$w_t$ & Wage rate
		& $e_t$ & Individual labor productivity levels \\
		
		$h_t$ & Working hours
		& $a_t$ & Household's asset (wealth) \\
		
		$a_0$ & Household's initial asset (wealth)
		& $r_{\text{interest}}$ & Interest rate \\
		
		$i_t$ & Household's income
		& $N$ & Number of households \\
		
		$\rho_e$ & Persistence
		& $u_t$ & Standard normal shocks \\
		
		$\beta$ & Discount factor
		& $\theta$ & Coefficient of relative risk aversion (CRRA) \\
		
		$\gamma$ & Inverse Frisch elasticity
		& $\tau_s$ & Consumption tax rate \\
		
		$T_{\text{inc}}(i_t)$ & Income taxes
		& $T_{\text{ast}}(a_t)$ & Asset taxes \\
		
		$\tau_i, \tau_a$ & Average level of marginal income and asset tax
		& $\xi_i, \xi_a$ & Slope of marginal income and asset tax schedule \\
		
		$\tau_k$ & Capital tax
		& $Y_t$ & Firm's output \\
		
		$K_t$ & Capital for production
		& $L_t$ & Labor for production \\
		
		$\alpha$ & Capital elasticity
		& $G_t$ & Government social spending \\
		
		$G_t$ & Government spending
		& $B_t$ & Government debt \\
		
		\bottomrule
	\end{tabular}
	}
    \label{tab:var}
\end{table}
\section{Additional Experiment Results}

\subsection{Within-Episode Action Dynamics}
\label{sec:within_episode_actions}

\begin{figure*}[t]
    \centering
    \includegraphics[width=\textwidth]{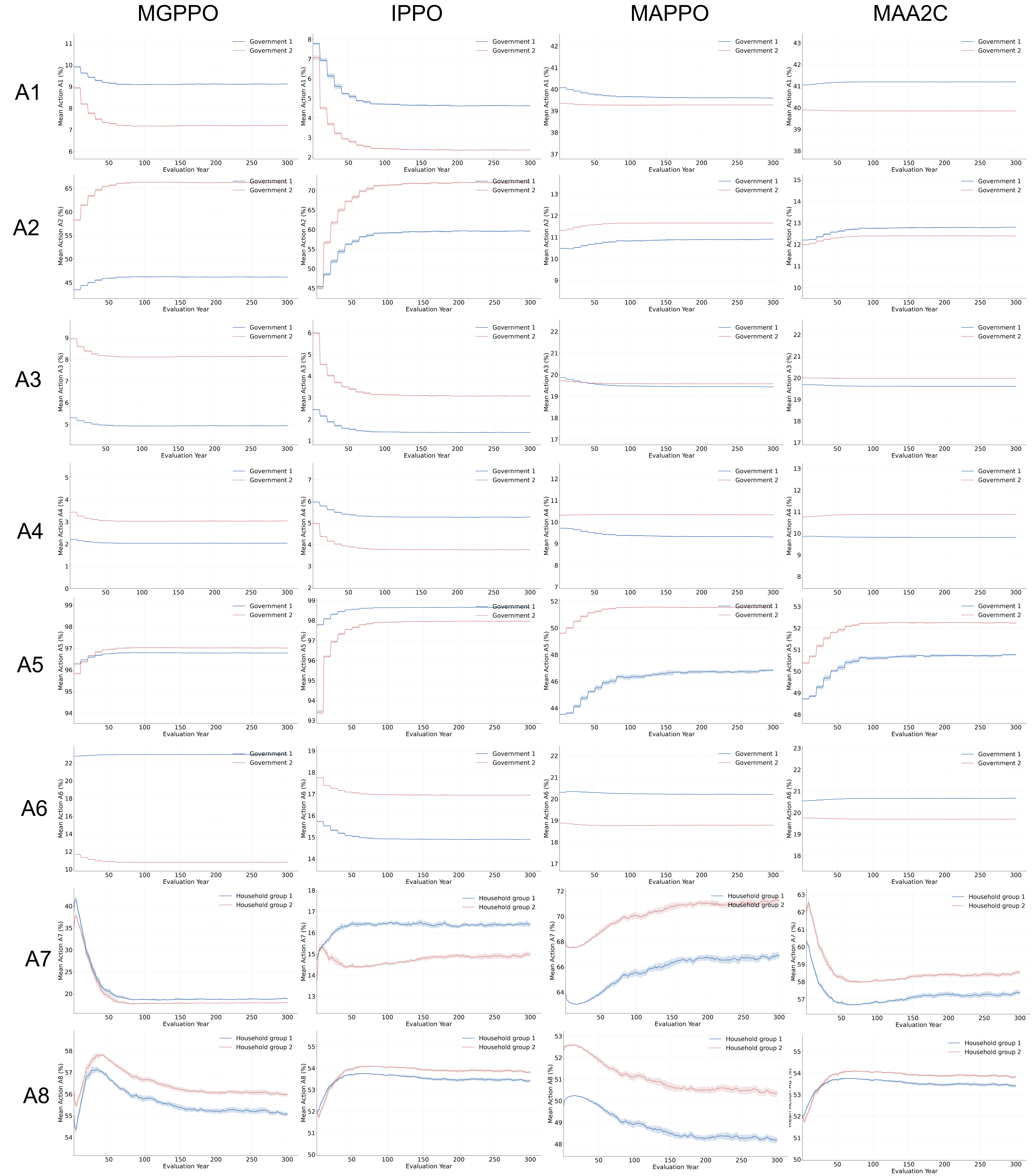}
    \caption{Within-episode action trajectories of trained IPPO, MAPPO, MAA2C, and MGPPO policies. A1--A6 denote the income-tax parameter, income-tax slope, asset-tax parameter, asset-tax slope, spending-ratio action, and capital-tax parameter, respectively. A7 and A8 denote the household saving-ratio and working-hour actions.}
    \label{fig:2.5}
\end{figure*}

To examine how trained policies respond to changing economic states, we evaluated each algorithm over a single episode in the two-group environment. Figure~\ref{fig:2.5} presents the resulting action trajectories. Government actions form plateaus because each choice remains fixed until the next government decision point, whereas households update their actions at every step.

Under MGPPO, the household saving action declined early in the episode before stabilizing, whereas the working-hour action changed little. The baseline algorithms followed distinct saving trajectories: IPPO remained comparatively stable, MAPPO increased, and MAA2C declined more gradually. Thus, MGPPO showed a larger within-episode adjustment in saving than in labor. In the household budget equation, saving allocates current resources between consumption and next-period assets, whereas working hours affect both labor income and effort costs.

The two MGPPO governments followed distinct policy trajectories, particularly for the income-tax slope and capital-tax actions. Their policies remained differentiated as economic states evolved. Differences in capital-tax actions represent one channel of modeled competition because capital flows depend on relative tax rates. Government actions under MAPPO and MAA2C varied less over the episode.

\section{Algorithm Implementation}
\label{app:alg}

Algorithm~\ref{alg:hierarchical_marl} summarizes the training procedure of the proposed multi-group bilevel learning framework. Training combines two-timescale action sampling, a capital-mobility curriculum, and on-policy updates.

At each environment step, households sample actions conditioned on their group-level and private observations and the current joint government action. Each government samples a new action once every $L$ steps and holds it fixed between sampling points. During the first $W$ episodes, government actions are selected deterministically while household policies continue to train.

At the beginning of episode $k$, the curriculum sets the capital-mobility weight $\phi_k$, which controls the strength of inter-group capital flows during trajectory collection. Household transitions are recorded at every environment step, whereas each government transition summarizes one action block using its discounted reward and the mean household action over that block.

After collecting $B$ episodes, the algorithm computes GAE-based $\lambda$-returns at the environment-step scale for households and at the action-block scale for governments. For each group, the critic is trained on these returns, while the actor is updated with clipped PPO using normalized returns as policy weights. Household networks are updated throughout training; government networks are updated only after the warm-up phase. The on-policy buffers are then cleared.

\begin{algorithm}[t]
\caption{Multi-Group Proximal Policy Optimization (MGPPO)}
\label{alg:hierarchical_marl}
\begin{algorithmic}[1]
\REQUIRE Number of groups $N$, training episodes $K$, maximum episode length $T$, government action interval $L$, episodes per update $B$, warm-up length $W$, curriculum horizon $M$, maximum capital-mobility weight $\phi_{\max}$, and government discount factor $\gamma_g$.
\STATE Initialize a government actor--critic and a household actor--critic for each group.
\STATE Initialize on-policy buffers $\mathcal D_g,\mathcal D_h\leftarrow\varnothing$ and buffered episode count $m\leftarrow0$.

\FOR{$k=0,\ldots,K-1$}
    \STATE Reset the environment and obtain initial observations.
    \STATE Set the previous joint government action $\mathbf a^g_{\mathrm{prev}}\leftarrow\mathbf0$ and block length $d\leftarrow0$.
    \STATE Set $\phi_k\leftarrow\phi_{\max}\min(1,k/M)$ if $M>0$; otherwise set $\phi_k\leftarrow\phi_{\max}$.

    \FOR{$t=0,\ldots,T-1$}
        \IF{$d=0$}
            \STATE Sample government actions using current observations and $\mathbf a^g_{\mathrm{prev}}$; select actions deterministically if $k<W$.
            \STATE Set block reward $\mathbf R^g\leftarrow\mathbf0$ and household-action sum $\mathbf S^h\leftarrow\mathbf0$.
        \ENDIF

        \STATE Sample household actions conditioned on group-level and private observations and the current joint government action.
        \STATE Compute the mean household action $\bar{\mathbf a}^h_t$ for each group.
        \STATE Execute the actions with capital-mobility weight $\phi_k$ and store household transitions in $\mathcal D_h$.
        \STATE Set $\mathbf R^g\leftarrow\mathbf R^g+\gamma_g^d\mathbf r^g_t$, $\mathbf S^h\leftarrow\mathbf S^h+\bar{\mathbf a}^h_t$, and $d\leftarrow d+1$.

        \IF{$d=L$ \OR the episode ends \OR $t=T-1$}
            \STATE Store a government transition in $\mathcal D_g$ with block reward $\mathbf R^g$, mean household action $\mathbf S^h/d$, and discount $\gamma_g^d$.
            \STATE Set $\mathbf a^g_{\mathrm{prev}}$ to the current joint government action and $d\leftarrow0$.
        \ENDIF
        \IF{the episode ends}
            \STATE \textbf{break}
        \ENDIF
        \STATE Advance to the next observations.
    \ENDFOR

    \STATE Set $m\leftarrow m+1$.
    \IF{$m=B$}
        \STATE Compute GAE-based $\lambda$-returns for household steps and government action blocks.
        \FOR{group $n=1,\ldots,N$}
            \STATE Update household actor--critic $n$ with clipped PPO, using normalized returns as policy weights.
            \IF{$k-B+1\geq W$}
                \STATE Update government actor--critic $n$ in the same manner.
            \ENDIF
        \ENDFOR
        \STATE Clear $\mathcal D_g,\mathcal D_h$ and set $m\leftarrow0$.
    \ENDIF
\ENDFOR
\end{algorithmic}
\end{algorithm}
\end{document}